\documentclass[english,aip,jcp,reprint]{revtex4-1}
\usepackage[T1]{fontenc}
\usepackage[latin9]{inputenc}
\usepackage{amsmath}
\usepackage{graphicx}
\usepackage{amssymb}
\usepackage{esint}
\usepackage[usenames]{color}
\usepackage{braket}

\begin{document}
\title{The effect of atomic-scale defects and dopants on graphene 
electronic structure} 

\author{Rocco Martinazzo}
\email{rocco.martinazzo@unimi.it}
\affiliation{Dipartimento di Chimica Fisica ed Elettrochimica, Università degli Studi di Milano, v. Golgi 19, 20133, Milan, Italy}
\affiliation{CIMaINa, Università degli Studi di Milano, v. Celoria 16, 20133, Milan, Italy}
\affiliation{Istituto di Scienze e Tecnologie Molecolari, v. Golgi 19, 20133, Milan, Italy}
\author{Simone Casolo}
\affiliation{Dipartimento di Chimica Fisica ed Elettrochimica, Università degli Studi di Milano, v. Golgi 19, 20133, Milan, Italy}
\author{Gian Franco Tantardini}
\affiliation{Dipartimento di Chimica Fisica ed Elettrochimica, Università degli Studi di Milano, v. Golgi 19, 20133, Milan, Italy}
\affiliation{CIMaINa, Università degli Studi di Milano, v. Celoria 16, 20133, Milan, Italy}
\affiliation{Istituto di Scienze e Tecnologie Molecolari, v. Golgi 19, 20133, Milan, Italy}

\begin{abstract}
Graphene, being one-atom thick, is extremely sensitive to the presence of adsorbed atoms and molecules and, more generally, to defects such as vacancies, holes and/or substitutional dopants. This property, apart from being directly usable in molecular sensor devices, can also be employed to tune graphene electronic properties. Here we briefly review the basic features of atomic-scale defects that can be useful for material design. After a brief introduction on isolated $p_z$ defects, we analyse the electronic structure of multiple defective graphene substrates, and show how to predict the presence of microscopically ordered magnetic structures. Subsequently, we analyse the more complicated situation where the electronic structure, as modified by the presence of some defects, affects chemical reactivity of the substrate towards adsorption (chemisorption) of atomic/molecular species, leading to preferential sticking on specific lattice positions. Then, we consider the reverse problem, that is how to use defects to engineer graphene electronic properties. In this context, we show that arranging defects to form honeycomb-shaped superlattices (what we may call "supergraphenes") a sizeable gap opens in the band structure and new Dirac cones are created right close to the gapped region. Similarly, we show that substitutional dopants such as group IIIA/VA elements may have gapped quasi-conical structures corresponding to massive Dirac carriers. All these possible structures might find important technological applications in the development of graphene-based logic transistors. 
\end{abstract}

\maketitle

\tableofcontents

\section{Introduction}

Graphene, thanks to its extraordinary electronic and mechanical properties, is a potential candidate for a number of applications. Being one-atom thick, it is extremely sensitive to the presence of adsorbed atoms and molecules (either physisorbed or chemisorbed on the surface) and, more generally, to defects such as vacancies, holes and/or substitutional dopants. This property, apart from being directly usable in molecular sensor devices, can also be employed to tune graphene electronic properties. \\ 
In this Chapter we review those basic features of atomic-scale defects that can be useful for material design. After a brief introduction (Section \ref{sec:2DEG}) of the main properties determining the peculiar electronic structure of graphene, and the experimental realisation of defective substrates (Section \ref{sec:exp}), we focus in Section \ref{sec:LD} on isolated ``p$_z$ defects'' such as atom vacancies or adsorbed species which covalently bind carbon atoms. In particular, we discuss in detail the formation of so-called midgap states and the microscopically ordered magnetic structures which give rise to. In Section \ref{sec:HD} we analyse the electronic structure of multiple defective graphene substrates and show, in particular, how it is possible to use simple rules to predict the presence of magnetic moments and midgap states by looking at the defect locations on the lattice.
Subsequently, we analyse the more complicated situation where the electronic structure, as modified by the presence of some defects, affects chemical reactivity of the substrate towards adsorption (chemisorption) of atomic/molecular species, leading to a preferential sticking on specific lattice positions. 
In Section \ref{sec:superlattices} we consider the reverse problem, that is how to use defects (vacancies, adsorbed species, substitutional dopants, etc..) to engineer graphene electronic properties. This is possible nowadays since recent advances in lithographic and self-assembling techniques allow one to produce well-ordered structures and thus 
`tune' the electronic bands. In this context, we show for instance how it is possible to open a band-gap in graphene and preserve at the same time the pseudo-relativistic behaviour of its charge carriers. We further analyse the case of substitutional dopants (group IIIA/VA elements) which, if periodically arranged, may show a gapped quasi-conical structure corresponding to massive Dirac carriers. All these possible structures might find important technological applications in the development of novel graphene-based logic transistors.

\section{The $\pi$-electron gas}\label{sec:2DEG}

Carbon atoms in graphene are arranged to form a honeycomb lattice tightly held 
by strong $\sigma$ bonds between $sp^2$ orbitals which form occupied $\sigma$ bands at energies well below the Fermi level.
The remaining valence electrons (one for each carbon atom) populate a $\pi$ band which localises above and below the lattice with a node on the surface plane. 
An `antibonding' $\pi^*$ band is empty when the system is at $T=0$ K and charge-neutral, but can easily be occupied, \emph{e.g.} by applying a gate potential in a typical field-effect transistor (FET) configuration. Such $\pi/\pi^*$ band system governs the low-energy (say up to $\sim2$ eV) behaviour of charge carriers in graphene and is responsible for most of the extraordinary properties of this material. This ``$\pi$ cloud'' is the focus of this section, where we introduce the main theoretical tools used in this Chapter. 

\subsection{Tight-binding Hamiltonian}

\begin{figure*}[t]	
  \centering
  \includegraphics[height=4.5cm]{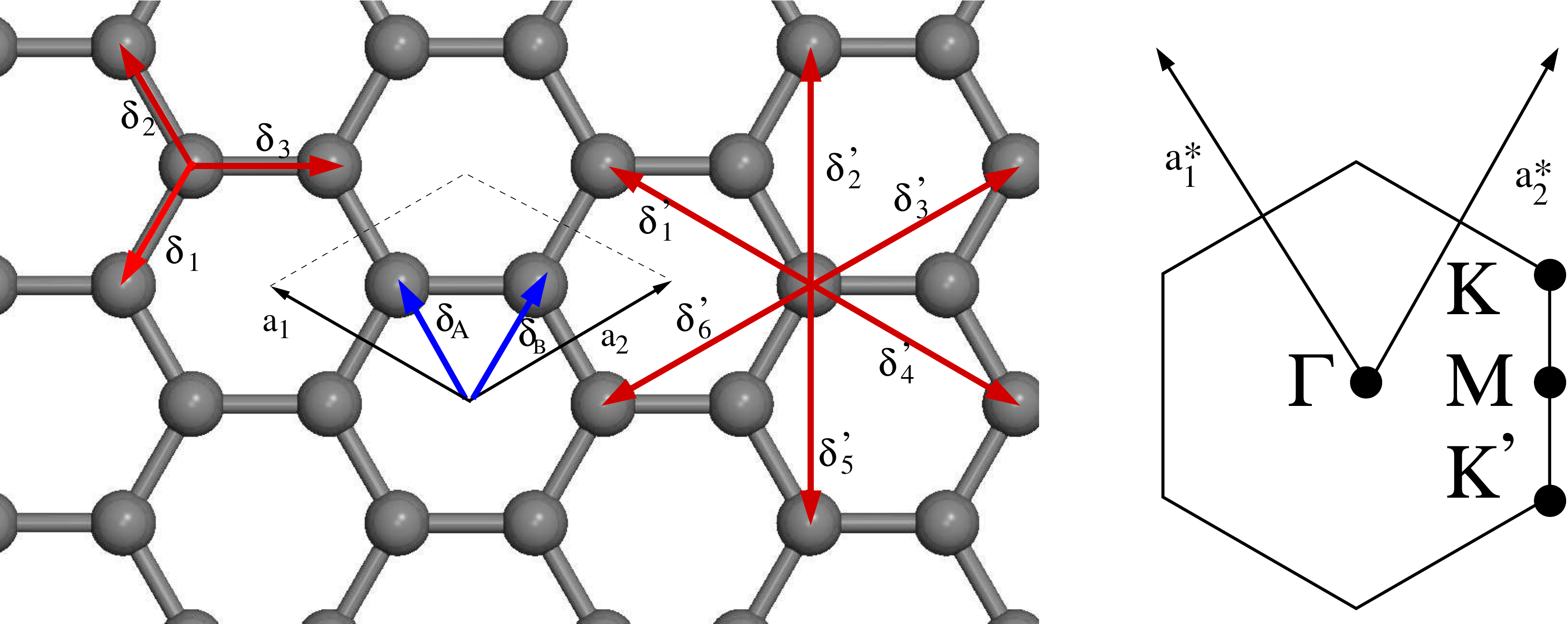}
  \caption{ Left panel: graphene unit cell ($\mathbf{a}_1,\mathbf{a}_2$), along with the vectors joining nearest- and next-to nearest neighbours, $\boldsymbol{\delta}_i$ and  $\boldsymbol{\delta}'_i$ respectively. Also indicated the position vectors $\boldsymbol{\delta}_A$ and $\boldsymbol{\delta}_B$ for A and B sites. The Wigner-Seitz cell is one the hexagons. Right: first Brillouin zone with the highest symmetry point indicated. The arrows are the reciprocal lattice vectors ($\mathbf{a}^{*}_1,\mathbf{a}^{*}_2$). }\label{fig:WS-BZ}
\end{figure*} 
In building up a simple, one-electron model for these $\pi$ electrons only one writes the one-electron wave function as a linear combination of two Wannier 
basis functions built with $p_z$ orbitals, one for each sublattice \citep{wallace47},
\begin{equation*}
\psi_{\mathbf{k}}({\mathbf{r}})=c_A\psi^A_{\mathbf{k}}({\mathbf{r}})+c_B\psi^B_{\mathbf{k}}({\mathbf{r}})
\end{equation*}
\begin{equation*}
\psi^A_{\mathbf{k}}({\mathbf{r}})=\frac{1}{\sqrt{N}}\sum_{j\in S}
e^{-i\mathbf{kr}} p_z({\mathbf{r}}-{\mathbf{R}}_{j}^A)
\end{equation*}
(and similarly for $\psi^B_{\bf{k}}({\bf{r}}) $) where the sum runs over lattice vectors
$\bf{R}_j$ within a large supercell \emph{S} including $N$ graphene unit cells 
and $\mathbf{R}_j^{A}=\mathbf{R}_j+\boldsymbol{\delta}_{A}$ is the position of $A$ site in the $j$-th cell
Equivalently, in second-quantized form 
\begin{eqnarray}
\hat{H}^{TB} & = & -t_{1}\sum_{<i,j>}\sum_{\sigma}\left(\hat{a}_{i,\sigma}^{\dagger}\hat{b}_{j,\sigma}+h.c.\right)+\nonumber \\
 &  & -t_{2}\sum_{\ll i,j\gg}\sum_{\sigma}\hat{a}_{i,\sigma}^{\dagger}\hat{a}_{j,\sigma}+\label{eq:TB}\\
 &  & -t_{2}\sum_{\ll i,j\gg}\sum_{\sigma}\hat{b}_{i,\sigma}^{\dagger}\hat{b}_{j,\sigma}+etc.\nonumber \end{eqnarray}
where $\hat{a}^{\dagger}_{i,\sigma}$ ($\hat{b}^{\dagger}_{i,\sigma}$) 
creates an electron with spin $\sigma=\uparrow , \downarrow$ on the 
$i$-th lattice site of the A(B) sublattice, the first two sums run over nearest neighbouring sites ($t_1$ is the hopping energy) 
and the second ones over sites which are nearest neighbours in each sublattice ($t_2$ is the corresponding hopping)
In absence of magnetic fields the hoppings can be chosen real, and the accepted value for $t_1$ is $\sim 2.7$ eV while $|t_2|<<t_1$ depends on the parametrization used. Neglecting overlap between orbitals on different C atoms, the usual anticommutation rules $[\hat{c}_{i,\sigma}^{\dagger},\hat{c'}_{j,\sigma '}]_{+}=\delta_{c,c'} \delta_{i,j}\delta_{\sigma , \sigma '}$ ($c=a,b$) hold; hence, introducing the Fourier transformed operators $\hat{a}_{\mathbf{k},\sigma}$ according to 
\begin{equation*}
\hat{a}_{i,\sigma}=\frac{1}{\sqrt{N}}\sum_{\mathbf{k}\in BZ}e^{-i\bf{k}\bf{R}_i}\hat{a}_{\mathbf{k},\sigma}
\end{equation*}
where the sum runs over $k$ points in the first Brillouin zone (BZ) (analogously for $\hat{b}_{\mathbf{k},\sigma}$) the above Hamiltonian can be rewritten as
\begin{eqnarray*}
\hat{H}^{TB} & = & -t_{1}\sum_{\mathbf{k},\sigma}f(\mathbf{k})\hat{a}_{\mathbf{k},\sigma}^{\dagger}\hat{b}_{\mathbf{k},\sigma}+h.c.\\
 &  & -t_{2}\sum_{\mathbf{k},\sigma}g(\mathbf{k})\hat{a}_{\mathbf{k},\sigma}^{\dagger}\hat{a}_{\mathbf{k},\sigma}-t_{2}\sum_{\mathbf{k},\sigma}g(\mathbf{k})\hat{b}_{\mathbf{k},\sigma}^{\dagger}\hat{b}_{\mathbf{k},\sigma}\end{eqnarray*}
or, in matrix notation, 
\[\hat{H}^{TB}=-\sum_{\mathbf{k},\sigma}\left[\hat{a}_{\mathbf{k},\sigma}^{\dagger},\hat{b}_{\mathbf{k},\sigma}^{\dagger}\right]\left[\begin{array}{cc}
            t_{2}g(\mathbf{k}) & t_{1}f(\mathbf{k})\\
            t_{1}f^{*}(\mathbf{k}) & t_{2}g(\mathbf{k})\end{array}\right]\left[\begin{array}{c}
            \hat{a}_{\mathbf{k},\sigma}\\
            \hat{b}_{\mathbf{k},\sigma}\end{array}\right]\]

Here $f(\bf{k})$  and $g(\bf{k})$  are `structure factors' for the nearest- and next-nearest neighbours,
\[f(\mathbf{k})=\sum_{i=1,3}e^{-i\mathbf{k}\boldsymbol{\delta}_i}\]
\[g(\mathbf{k})=\sum_{i=1,6}e^{-i\mathbf{k}\boldsymbol{\delta}^{'}_i}\]
Diagonalization is trivial and gives the energy bands, 
\begin{equation}
\epsilon({\bf{k}})_{\pm}=-t_2g(\mathbf{k})\pm t_1 |f(\mathbf{k})|=-t_2g(\mathbf{k})\pm t_1 \sqrt{3+g(\mathbf{k})} 
\label{eq:bands}
\end{equation}
where $|f(\mathbf{k})|^2=3+g(\mathbf{k})$ has been used and the minus (plus) sign solution correspond to the $\pi$ ($\pi^*$) band 
(see \emph{e.g.} Ref.s \onlinecite{wallace47},\onlinecite{castroneto09},\onlinecite{Bena09}).
Close to the $K$($K'$) point $|f(\mathbf{K+q})|^2\sim v_F^2q^2$ and the dispersion is conical, giving rise to the so-called Dirac cones. Here $v_F=\frac{\sqrt{3}}{2}t_1 a=\frac{3}{2}t_1 d$, where $d$ is the carbon-carbon distance, $\sim1.42$ \AA,  and $a$ the lattice constant. Consequently, the density-of-states (DOS) is linearly vanishing at zero energy, $\rho(\epsilon)\sim4|\epsilon|/\pi\sqrt{3}t_1^2$ (per cell, with spin and valley degeneracies included), one of the fingerprints of massless Dirac electrons. Its vanishing value challenges one's intuition since experiments find a finite, non-zero minimum conductivity at this energy \citep{peres10}.\\
\begin{figure*}[t]	
  \centering
  \includegraphics[height=4cm]{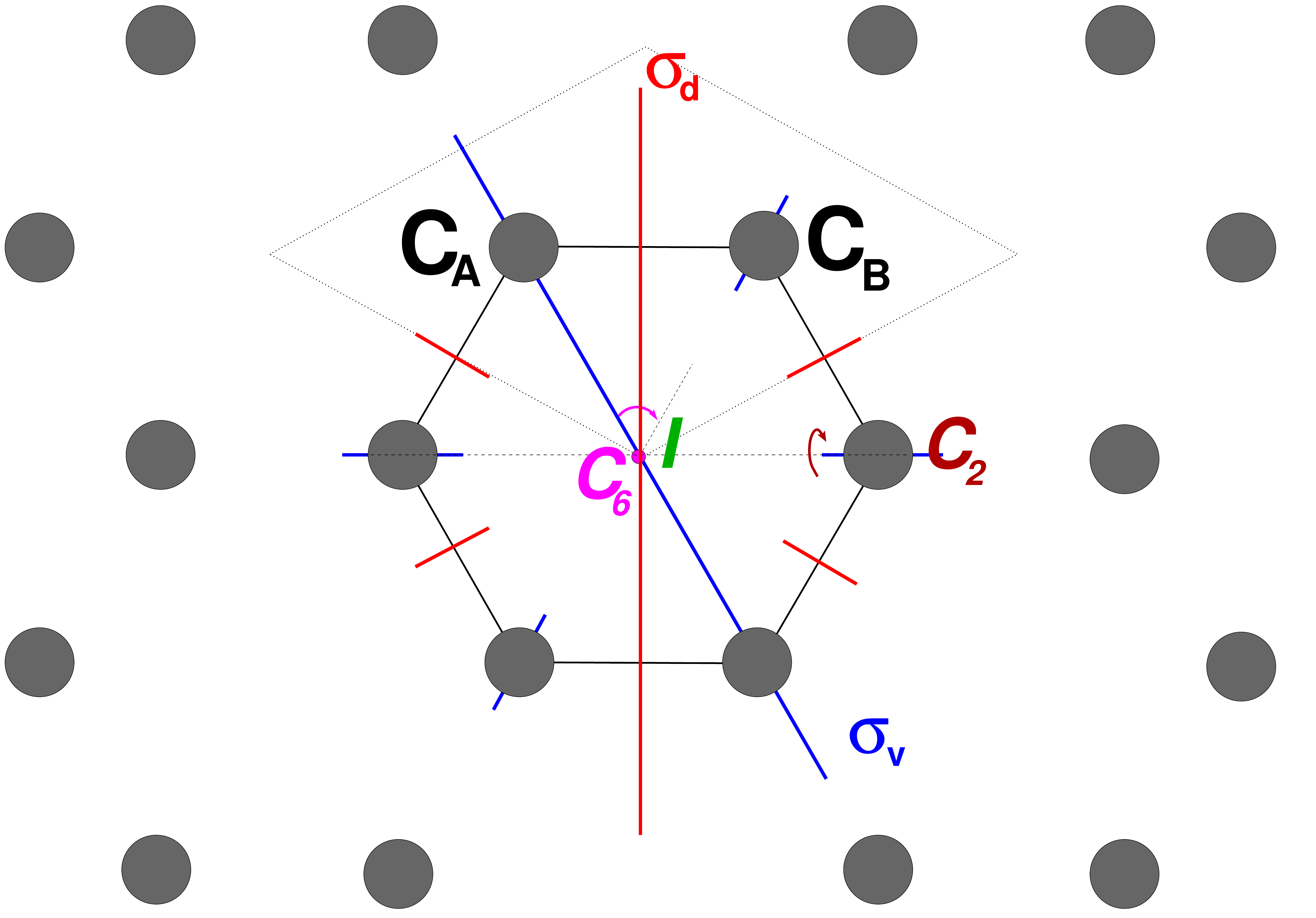}
  \includegraphics[height=4cm]{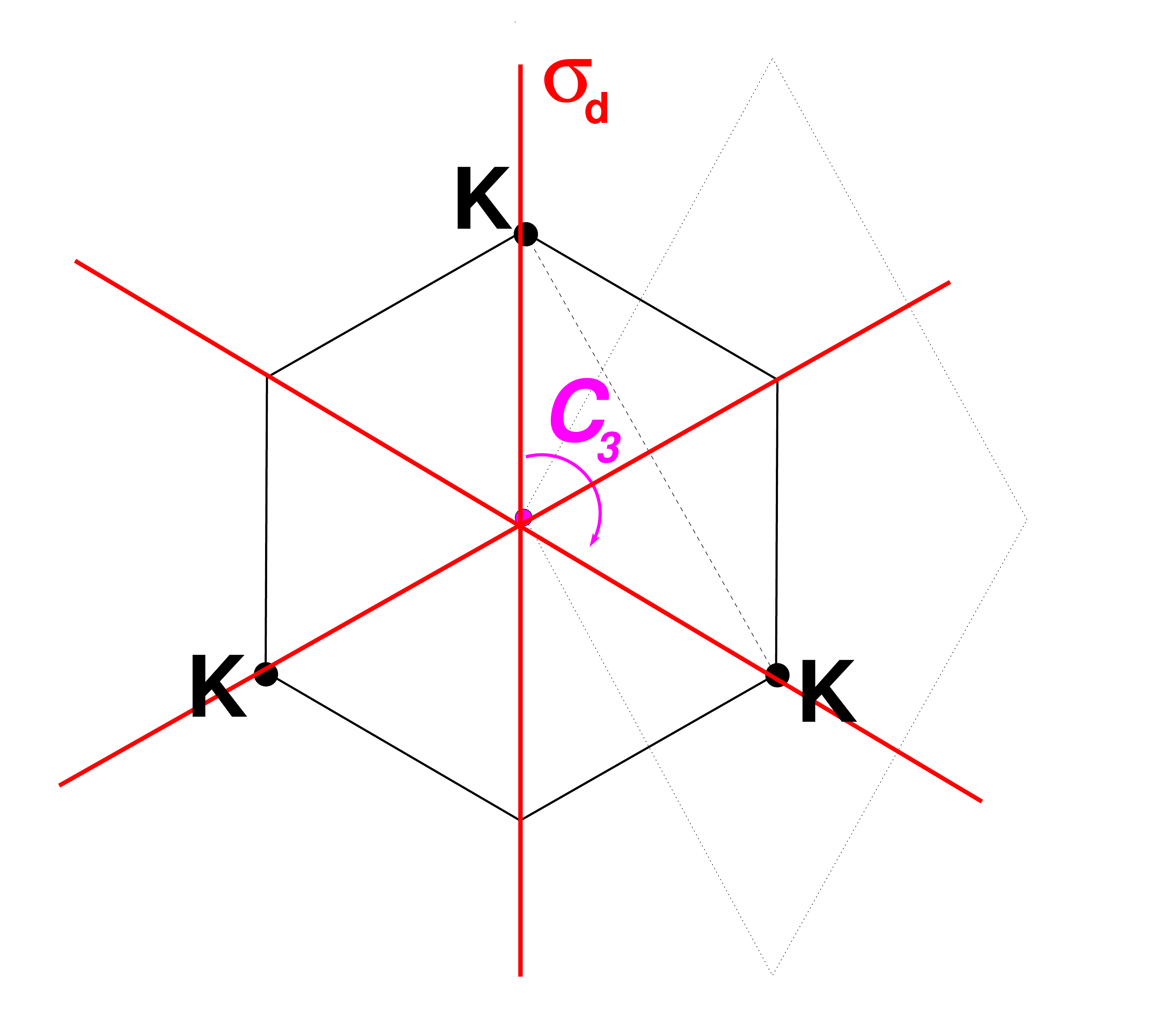}
  \caption{ Left panel: Point symmetry elements in graphene lattice. Reflection planes orthogonal to the page, $\sigma_v$ and $\sigma_d$, are replicated by the six-fold rotation axis $C_6$, along with the two-fold rotation axis on the page plane, $C_2$. $I$ is the inversion center, at the center of the Wigner-Seitz cell (solid line). The page plane is, of course, a reflection plane ($\sigma_h$). Right panel: symmetry elements of the $k$-group at the $K$ point ($D_{3h}$). Black dots mark the $K$ point and its images.  }\label{fig:symm}
\end{figure*}   
Albeit simple, this tight-binding model is accurate enough to correctly represent graphene $\pi$ bands, at least close to the high 
symmetry points  $K$ and  $K'$. The latter control the low-energy physics of charge carriers, and are the source of the exceptional interest in graphene.  
If only nearest-neighbours interaction is allowed the two sublattices form two disjoint sets where $A$-type sites connect to $B$-type sites only and \emph{vice versa}. 
The Hamiltonians is said \emph{bipartitic} and displays an interesting symmetry: for each non-zero energy level $\epsilon$ and eigenfunction $\ket{\psi_{+}}=c_{A}\ket{A}+c_{B}\ket{B}$ (where $\ket{A}$/$\ket{B}$ is non-zero on $A$/$B$ lattice sites only), there exists a `conjugate' level with energy $-\epsilon$ and wavefunction   $\ket{\psi_{-}}=c_{A}\ket{A}-c_{B}\ket{B}$. This is called \emph{electron-hole} ($e-h$) symmetry since at \emph{half-filling} (as it is case of graphene with one electron per site), the Fermi level lies at zero energy, and the above symmetry relates electron and holes. For a proof, just apply a phase-change to one of the two sets of sublattices states \footnote{Interestingly, this operation corresponds to an operator $\hat{\pi}$, $\hat{\pi} \hat{c}_{i,\sigma}=(-)^\tau\hat{c}_{i,\sigma}\hat{\pi}$ ($\tau=1,2$ for $c=a,b$), which reduces to the $z$ component of the pseudospin in spinor notation.}, \emph{e.g.} $\hat{b}_{i,\sigma} \rightarrow -\hat{b}_{i,\sigma}$, as this converts $\hat{H}$ into $-\hat{H}$.\\    
\begin{figure*}[t]	
  \centering
  \includegraphics[height=5cm]{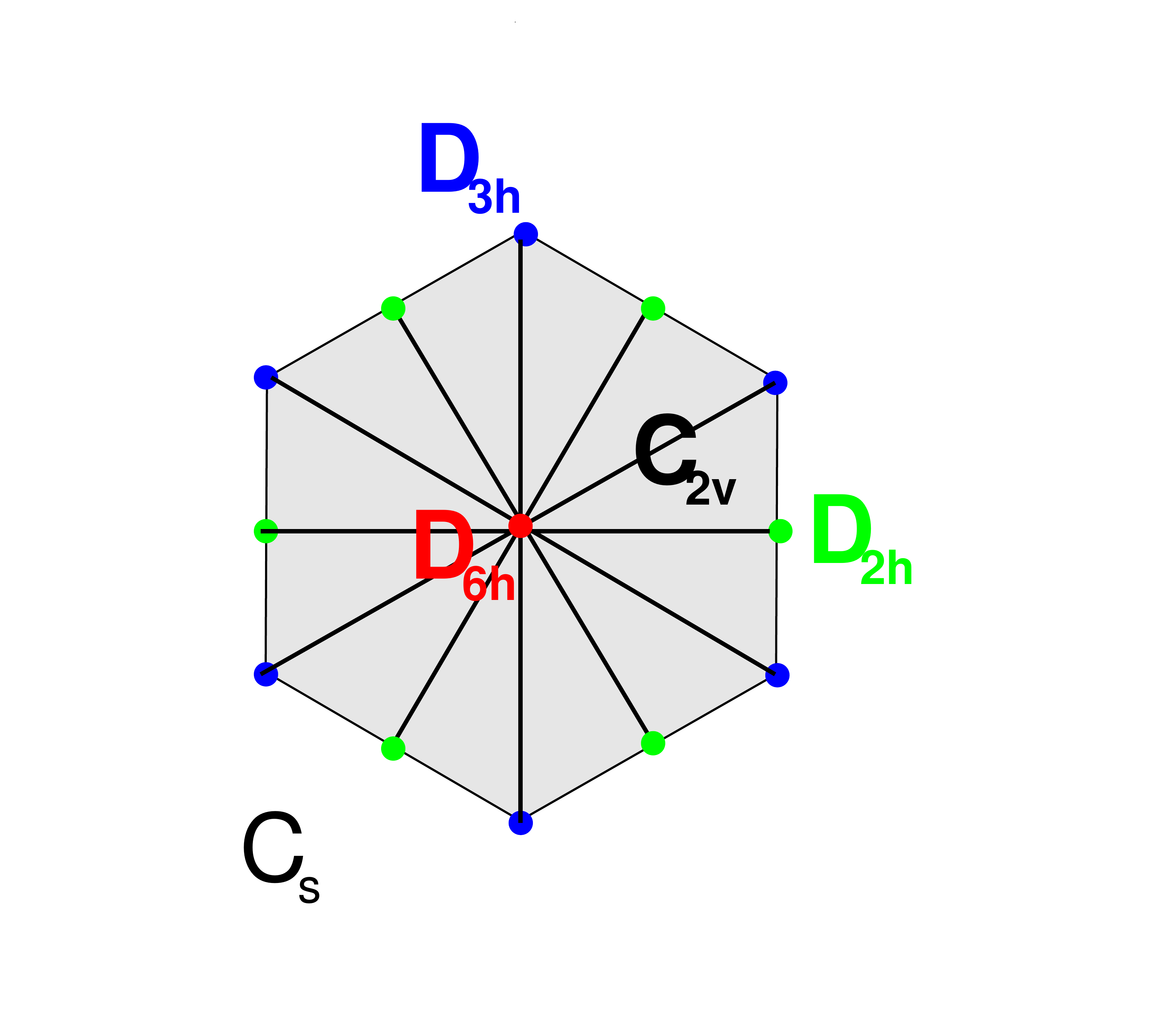}
  \includegraphics[height=5cm]{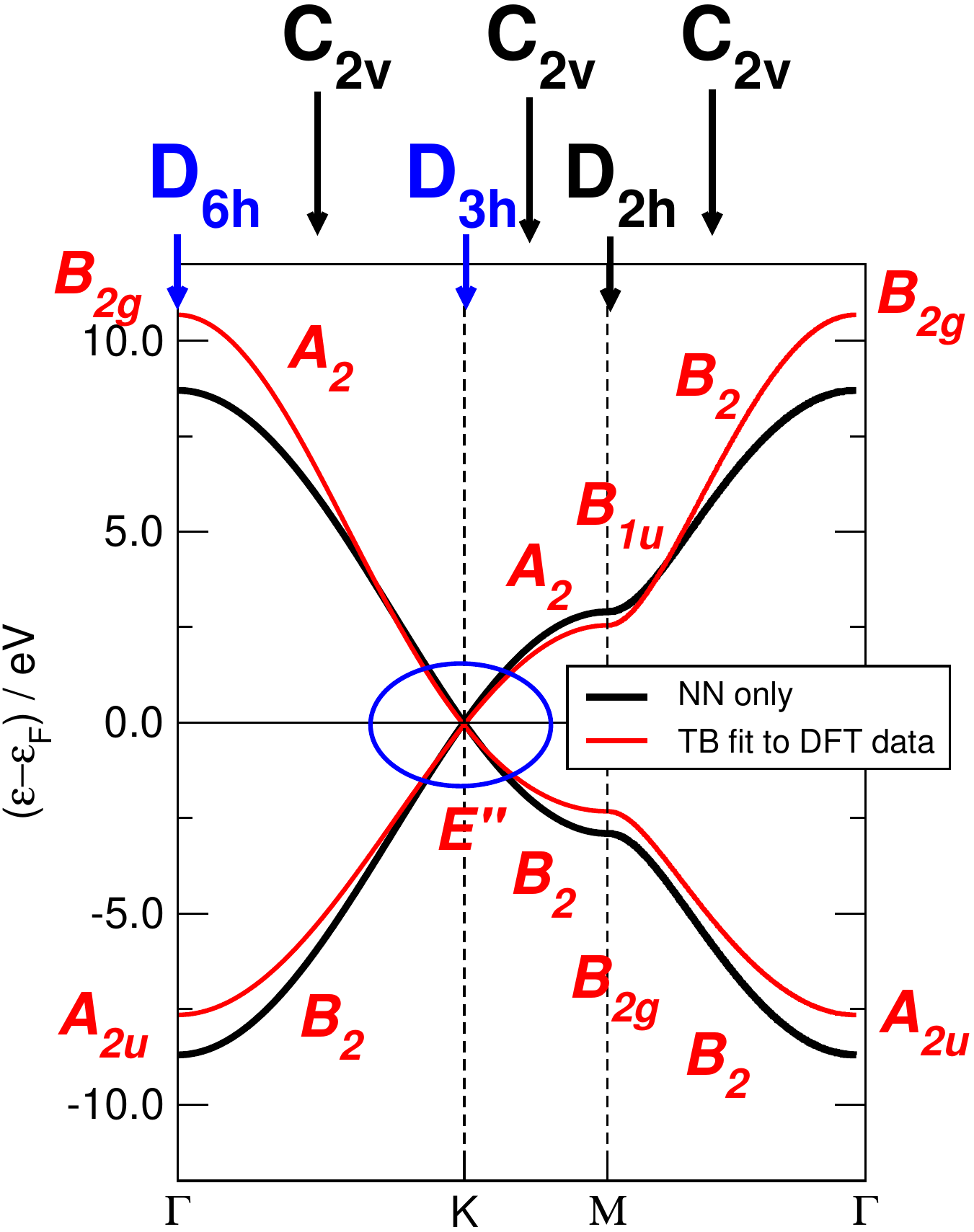}
  \caption{ Left panel: point-symmetry classification of Bloch states in graphene. $k$-groups are color coded as indicated. Right panel: symmetry labels for graphene electronic states on the highest-symmetry lines of the BZ. }\label{fig:symm2}
\end{figure*}  
Electron-hole symmetry, as we shall see in the following, plays an important role in graphene, even if it holds only approximately (\emph{i.e.} with nearest-neighbour interaction only and neglecting orbital overlap). Here we just notice that, because of such symmetry, the band-structure is expected to have a gap at the Fermi level unless there are specific reasons for having energy levels exactly at $zero$.
As we now show, the specific reason is provided by the \emph{spatial} symmetry of the substrate.\\   
Graphene lattice is highly symmetric. Its Wigner-Seitz unit cell has the same point symmetry of benzene, namely it belongs to the $D_{6h}$ point group, see fig.\ref{fig:symm}, which is the point group for symmetry operations in real-space. For Bloch electronic states with $k$-vector $\mathbf{k}$, symmetry is reduced to that subgroup of $D_{6h}$ which either leaves $\mathbf{k}$ invariant or transform it into one of its images, \emph{i.e.} $\mathbf{k}\rightarrow \mathbf{k}+\mathbf{G}$ with $\mathbf{G}$ a reciprocal lattice vector \citep{mirman}. Such subgroup is known as $k$-group at $\mathbf{k}$, $G(\mathbf{k})$, and determines the possible symmetry of the electronic states. For instance, at the $K$ point the $k$-group is $D_{3h}$ since only three-fold rotation axes and $\sigma_d$ planes transform the $K$ images into themselves \footnote{The remaining symmetry elements determine the so-called \emph{star} of the given $k$ point, which is the set of points generated by these elements once applied to ${\bf k}$. Such set of physically distinct points in $k$ space are degenerate in energy; this is the case of $K'$ and $K$, for instance, since they belong to the star of each other.}.  A full analysis of the symmetry properties of Bloch electrons is given in fig.\ref{fig:symm2}, left panel, where the $k$-groups are color-coded, grey for $C_s$, black for $C_{2v}$, green for $D_{2h}$, blue for $D_{3h}$ and red for $D_{6h}$. The main point here is that graphene is sufficiently symmetric that allows $k$-groups supporting $two$-dimensional irreducible representations ($E$ irreps), namely $D_{6h}$ at $\Gamma$ and $D_{3h}$ at $K,K'$. As spatial symmetry is (almost) compatible with $e-h$ symmetry, a zero energy state results whenever the electronic wavefunctions span a two-dimensional irreducible representation (odd in number, in general), \emph{i.e.} they give rise to a doubly-degenerate level. This is exactly the case of the $K$ ($K'$) point, where Wannier functions built with $p_z$ orbitals of the A and B sublattice span the $E''$ irrep of the above $D_{3h}$ $k$-group
. Notice also that this symmetry argument is enough to explain the conical dispersion of the energy at the $K$ ($K'$) point which makes graphene so attractive: without an inversion center, degeneracy is lifted already at first order in $\mathbf{k\cdot p}$ perturbation theory when moving away from the BZ corners.\\ 
While spatial symmetry is exact, $e-h$ symmetry holds in the nearest-neighbour approximation only. Nevertheless, since inclusion of higher order hopping terms does not modify the level ordering (\emph{i.e.} the minimum of the $\pi^*$ band lies always above the top of the $\pi$ band) the Fermi level at charge 
neutrality matches exactly the energy where the $E$ irrep is found (called Dirac point
 as it is the cone apex).

\subsection{Hubbard Hamiltonian}

The tight-binding (TB) Hamiltonian is a model in which each electron moves independently 
from the others. Despite it represents a good approximation for graphene 
energy spectrum, such a simple picture will necessarily fail in computing 
spin properties in all but the simplest situations. A simple way to include 
electron-electron interactions is given by the Hubbard model
\begin{equation}
\hat{H}=\hat{H}^{TB}  + U \sum_{i}\hat{n}_{i,\uparrow}\hat{n}_{i,\downarrow}
\end{equation}
where $\hat{H}^{TB}$ is the tight-binding Hamiltonian of eq.\ref{eq:TB}, the sum runs over all carbon sites and 
$\hat{n}_{i,\sigma}=\hat{c}^{\dagger}_{i,\sigma}\hat{c}_{i,\sigma}$   
are the corresponding number operators.
This Hamiltonian combines the tendency of electrons to delocalize onto the 
lattice due to their kinetic (hopping) energy together with an ``on-site'' Coulomb
repulsion that tends (for $U>0$) to localize them to minimize double orbital occupation.\\
The Hubbard model is a very useful tool for the study of magnetism in 
complex materials. It has long been used in the chemical community\footnote{In the chemical community is dubbed Parisier-Parr-Pople approximation, after Pariser-Parr and Pople who first introduced it in the early fifties.} -and proved to be rather accurate for such systems- 
to investigate excitation spectra in polycyclic aromatic hydrocarbons (today, graphene dots). 
Though simple, the model requires quite a large effort for its solution. Therefore, one often resorts to its mean-field approximation,
\begin{eqnarray*}
\hat{H}^{mf} & = & \hat{H}^{TB}+U\sum_{i}\hat{n}_{i,\uparrow}\braket{\hat{n}_{i,\downarrow}}+U\sum_{i}\braket{\hat{n}_{i,\uparrow}}\hat{n}_{i,\downarrow}\\
 &  & -U\sum_{i}\braket{\hat{n}_{i,\uparrow}}\braket{\hat{n}_{i,\downarrow}}\end{eqnarray*}
where the average occupation number of one spin-species at a given site tunes an effective on-site energy for the other spin-species, \emph{e.g.}  $\epsilon_{i,\uparrow}^{eff}=+U \braket{\hat{n}_{i,\downarrow}}$. This is essentially equivalent to an (unrestricted) Hartree-Fock approach to the $\pi$ electrons and is useful, as compared with density-functional-theory (DFT) methods applied to the exact Hamiltonian, to study very large systems, of dimension comparable to those experimentally realized. Though we will not solve the Hubbard model in the following, there are some exact, analytic results that can be obtained from it and that turn out to be important tools in discussing defects in graphene.

\subsection{Valence Bond picture}\label{sec:VB2}

\begin{figure}[t]
  \centering
      \includegraphics[width=0.9\columnwidth]{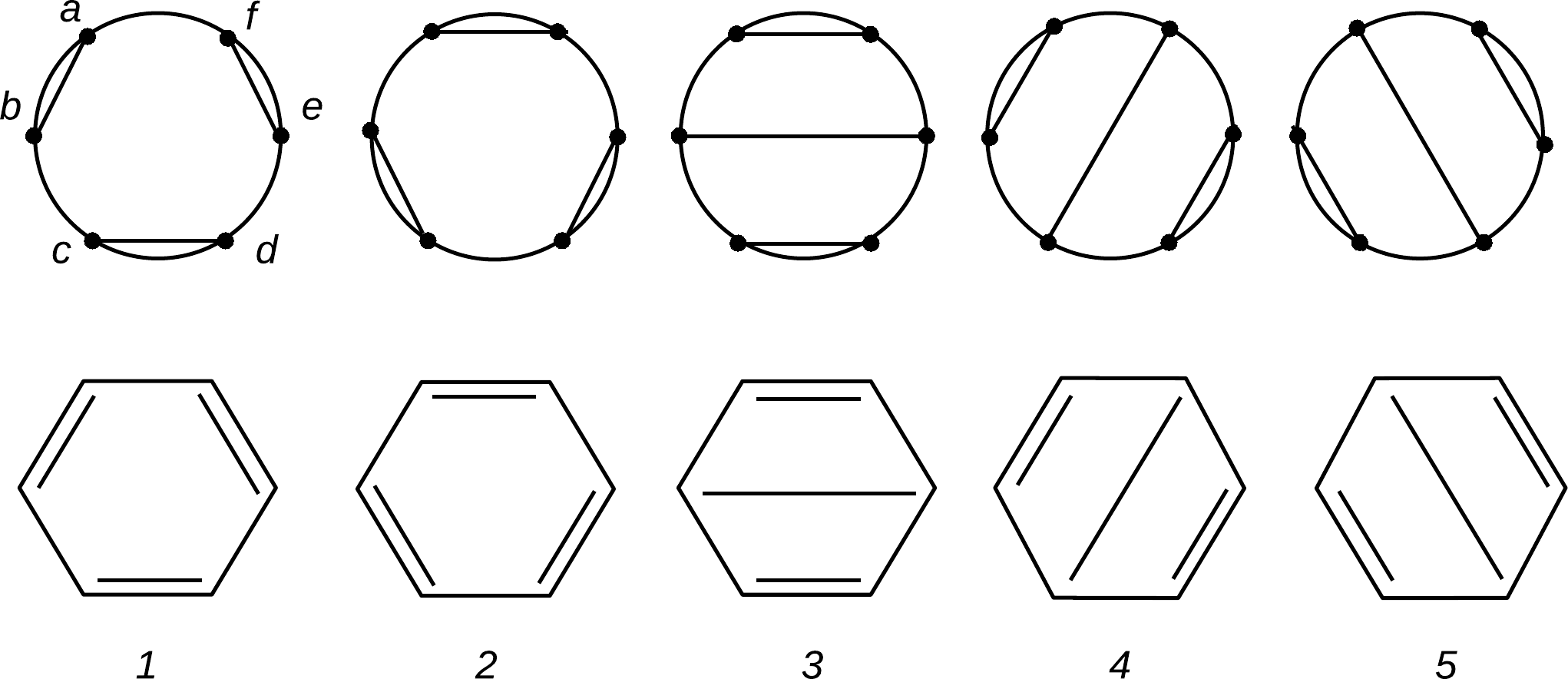}
  \caption{ \label{fig:rumer}The five possible perfect pairing
 Rumer diagrams for the  benzene molecule with their correspondence with Kekul\'e (1-2) and Dewar
  (3-5) resonance structures.}
\end{figure}

An alternative, easy-to-use way of looking at graphene electronic structure is provided by the `chemical picture'. 
With this we mean the traditional picture of chemical bonds as given by the Lewis structures and modified to account for the `chemical resonance'.   
In this picture, electrons are mostly localized in atomic orbitals (usually hybridized) of the atoms forming the molecule, 
and couple in singlet pairs to form bonds and lone-pairs. For carbon atoms in graphene the three $sp^2$ orbitals 
(with one electron each) are singlet-coupled with electrons in $sp^2$ orbitals of neighbouring sites.
The remaining electron (the one described by the TB Hamiltonian above) can couple with its counterpart of one of the three neighbours.
The state of the system is a superposition of these different ways of binding, and the system gains energy from such a \emph{resonance} phenomenon.\\ 
This na\"ive picture finds its root in the \emph{Valence Bond} (VB) theory of chemical bond, 
which developed from the Heitler-London study of the $H_2$ molecule, soon after the foundation of quantum mechanics.
The theory, as intensively pushed forward by Slater and Pauling, 
is a practical way of looking at the chemical bond and at the bond-breaking, bond-forming processes which are essential for chemical reactivity.
It can also be turned into a variational method for the many-electron problem 
which uses a correlated wavefunction \emph{ansatz} and captures the important part of the electron correlation 
\footnote{This is so because even the simplest VB wavefunctions can be re-written as linear combinations of Slater determinants, and include the so-called `static' correlation. The latter is essential for describing bond formation and near-degeneracies; in extended systems is responsible for Mott transitions.}\citep{tantardini85,Cooper87,cooperbook02,VBBook}. 
In many respects, it has to be considered complementary to the \emph{Molecular Orbital} (MO) approach, though the latter   
proved to be numerically more efficient. \\
Valence Bond theory focuses on spin and builds the singlet wavefunction of an even-numbered 
ground-state molecule as a `product' of singlet pairs, one for each bond (pairs of orbitals), thereby identifying a chemical formula.
For less standard species such as graphene, different products are equally likely and the correct wavefunction is the linear combinations of all the possible structures.   
For instance, let us look at
 the benzene molecule as prototypical case of aromatic compounds. 
Considering only the six $\pi$ electrons localized in their respective $p_z$ orbitals, the
possible linearly independent (``perfect pairing'') functions can be schematically depicted as in figure 
\ref{fig:rumer}; for six electron and an overall singlet state there are five couplings\footnote{The number of linearly independent spin-function for $N$ electrons in the $S$ spin state, usually denoted as $f_S^N$, can be easily obtained by angular momentum coupling rules. The properties of the corresponding spin spaces stem from their deep connection to the group of permutations of $N$ objects.}.
In the graphical representation of Fig.\ref{fig:rumer}, known as Rumer
diagrams, one represents each atomic center with a dot and uses a line for 
singlet coupling between them. The chemical picture (bottom row of the same figure) uses only
the 2 Kekul\'e structures on the left, since the other 3 so-called Dewar structures gives a negligible 
contribution to the energy, as can be guessed from the bond pattern.\\ 
Notice that VB theory is closely related to the Hubbard model discussed in the previous section: the atomic orbitals housing the electrons need not be those of the free atoms. If they are 'polarized' by the environment (\emph{e.g.} they are linear combinations of free-atomic orbitals) the VB \emph{ansatz} accounts both for localization and band-like behaviour, as in the Hubbard model. It is not hard to show, indeed, that the Hubbard model for the $H_2$ molecule can be obtained from a simple\footnote{It is enough to consider the so-called Coulson-Fisher wavefunction.} VB \emph{ansatz} to the two-electron wavefunction. In the following we will make a qualitative use of this chemical picture, as it provides insights into defect formation and properties; see also Ref. \onlinecite{clar10} for its role in interpreting details of STM pictures.

\section{Defect formation} \label{sec:exp}

In this Section we describe defect formation\footnote{With this we mean point defects, \emph{i.e.} adatoms, substituted or displaced atoms. It should be noticed, however, that other defects may also be important for the transport properties. }. As we shall see in the following, even though vacancies and adatoms turn out to have very similar effects on the $\pi$ electron system, we distinguish them here according to their preparation methods, \emph{i.e.} high ($e^-$, ions, etc.) \emph{vs.} low (neutrals) energy beams. The reason is that only in the first case defects can be considered randomly arranged. Adatoms at all but very low concentration tend to cluster on the surface, and understanding this phenomenon requires knowledge of how the electronic structure of the substrate is modified upon formation of the very first defects. \\
Notice, however, that defects are also naturally present in graphene as in any common material. Adatoms as hydrogen or hydrocarbons, for instance, can be introduced by the preparation method, while other point defects affecting transport properties, \emph{e.g.} charge-impurities, local potentials, etc., may result from imperfections on the substrates where graphene is accommodated.     

\subsection{Electron and ion bombardment}

The irradiation by high energy particles is the main tool for creating defects  
in graphene and in other carbon nanostructures.
When the projectile particle impinges on the structure it transfers energy to the 
lattice. In bulk materials (\emph{e.g.} in graphite) energy 
dissipation is rather effective, up to eventually stop the projectile, 
and it occurs through \emph{nuclear} and \emph{electronic} stopping mechanisms.
Nuclear stopping is due to the collisions between the projectile and the 
carbon nuclei, an essentially classical process governed by momentum transfer and Coulomb 
interaction. On the other hand, (inelastic) electronic stopping occurs by the 
many possible electron transitions in the material, hence 
promotion into conduction band (hot electrons), ionizations, but also through 
plasmon excitations, photoemissions, etc.
The relative importance of the two mechanisms depends on the beam 
energy, on the projectile mass and on the electronic structure of the 
target material. Nevertheless a microscopic theory of energy dissipation 
in nanostructures is still under study 
since the models developed for bulk materials cannot be easily applied in 
a reduced dimensionality material such as graphene \citep{KrasheninnikovBigRev}.\\
The mechanism for the defect formation has been studied intensively in the last decade.
In brief, when the energy transferred to an atom 
is larger than the so-called displacement 
threshold ($\sim$20 eV in case of graphite) this can leave its equilibrium 
position and move trough the bulk to form, for instance, a Frenkel pair 
or, for single layer graphene, a vacancy.
Large ions can produce multiple vacancies up to small holes in the lattice
depending on their size.
Electron beams produced in 
transmission electron microscopes (TEM) can instead be focused down to 
scales comparable
 to the carbon-carbon distance, giving a precise 
control of the induced damage up to form single vacancies. 
Moreover TEMs allow a real-time imaging of the 
damage process and of the chemical reaction that follows the vacancies 
formation \citep{Zettl08,Meyer2010}.\\
The formation of a single vacancy in graphene leaves three $\sigma$ 
dangling bonds and it removes a $\pi$ electron.
The first span a low-energy, one-dimensional irreducible representation ($A$ irrep in the following) of the (local) $D_{3h}$
point group and an $E$ irrep. Therefore, the ground-state is degenerate and undergoes 
a Jahn-Teller distortion: the closure of two dangling 
bonds to form a pentagon, with an energy gain of about 0.2 eV.
The strain induced by the other hexagons in the lattice prevents further 
distortions of the third unsaturated atom out of plane \citep{ElBarbary03} 
and the final magnetic moment for such a structure has been reported to be 
between 1.0 and 1.5 $\mu_B$ \citep{Yazyev2007,Lethinen2004}, localized on the 
unpaired site. When exposed to a hydrogen flux, the vacancy 
rapidly saturates its dangling bonds, with H atoms pointing slightly out
of the graphene plane \citep{Lethinen2004}.\\ 
In the
 case of neutral-atom bombardment, the projectile can also react to form a 
covalent bond with a carbon atom. This is what happens by irradiating samples with  
low energies hydrogen atoms. 
It has been shown that at very low 
densities the chemisorbed H atom defect produces STM images very similar
to the single vacancy case. As already mentioned, at higher densities H atoms tend instead 
to cluster in dimers or larger structures due to electronic effects that will be discussed in the following.
Nevertheless, when considering $\pi$ electrons only, vacancy and singly-bond 
chemisorbed species are equivalent, since a single electron is removed 
from the aromatic network of graphene. As an example, chemisorption of a single H atom is detailed in the following section.

\subsection{Sticking of atomic and molecular species}

A hydrogen atom impinging on graphene with a low collision energy can either
physisorb or chemisorb. The physisorption regime has long been probed with the help of 
selective-adsorption resonances in H atom scattering off graphite \citep{Ghio1980}. 
The extrapolated value for the physisorption binding energy ($\sim$40 meV)
to a single layer is in very good agreement with recent theoretical studies \citep{Bonfanti2006}. 
Physisorbed species are highly mobile and easily desorb from the surface since they couple only weakly with the substrate.    
For this reason, chemisorption turns out to be more interesting for graphene electronic structure engineering.\\
\begin{figure*}[t]
  \centering
      \includegraphics[width=0.75\textwidth]{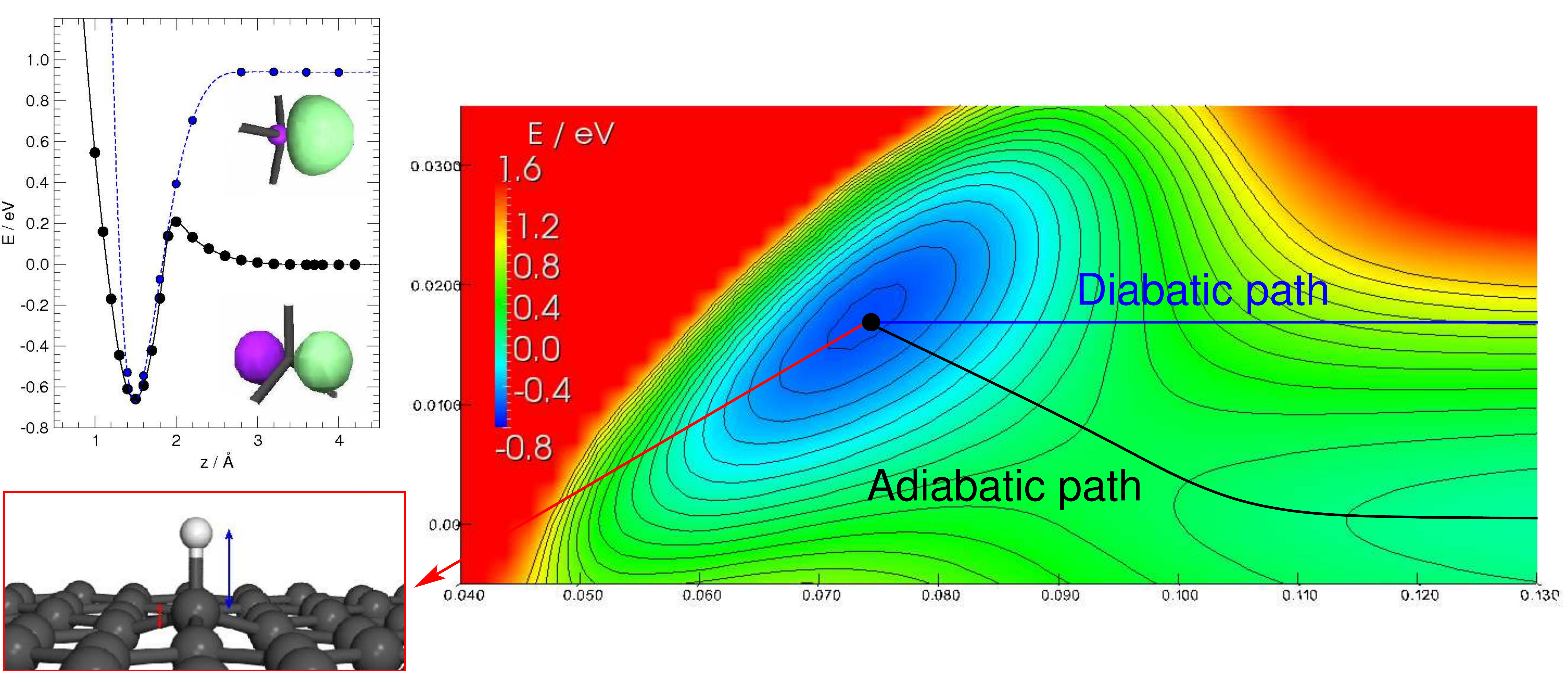}
      \caption{\label{fig:Hpath}Left Panel: chemisorption curves for a H atom on
        a graphene $top$ site as obtained from periodic DFT calculations using a 2x2 supercell. In the adiabatic path (black) the carbon atom is
        allowed to relax at each C-H distance, while in the diabatic path (blue) the top site is kept fixed in the puckered
        geometry. $z$ is the adsorbate height on the surface.
        The inset on the bottom shows the equilibrium geometry.
        Right panel: map of potential energy surface as a function of the heights of H and C atoms above the surface, for a collinear geometry. Also indicated the diabatic and adiabatic paths. Units are eV and \AA.}
\end{figure*}
Chemisorption of single H atoms on \emph{graphite} 
has been studied since the theoretical works of Jeloaica and Sidis\cite{Jeloaica1999} and Sha and Jackson\cite{Sha&Jackson2002}, who first showed 
that it indeed occurs if the substrate is allowed to relax.
Among the four possible adsorption sites the \emph{hollow} and \emph{bridge} 
were found not binding while the two kinds of \emph{atop} sites (with or without an carbon atom on the layer
underneath for graphite) give essentially the same behaviour, since graphene layers in graphite lay 
$\sim3.4$ \AA~apart.
This implies that the (surface) chemistry of graphene is very similar to that of graphite.
\\
Adsorption on the top site induces a surface reconstruction (`puckering').
Such a reconstruction consists in the outward motion of
the carbon atom beneath the adsorbed hydrogen, and occurs
as a consequence of $sp^2 - sp^3$ re-hybridization of the carbon valence
orbitals needed to form the CH bond. The re-hybridization induces a change 
in geometry of the substrate site, from a planar  ($sp^2$) to a tetrahedral ($ sp^3$) form,
thereby leading to the surface puckering. The energy required for such a process, defined as the energy difference between the relaxed and the puckered configuration, is substantial ($\sim 0.8$ eV)
 and this explains why binding energies to graphene are typically smaller than for other carbon species.  
If the graphene layer is kept flat the carbon - hydrogen bond is metastable only
 \citep{Sha&Jackson2002,casoloSIE},
while allowing surface relaxation chemisorption becomes an activated
process with stable products ($\sim 0.80$ eV).\\
When the hydrogen atom collides on a carbon site already
puckered, \emph{i.e.}
 already in the $sp^3$ form,
chemisorption is a barrierless process.
Otherwise, following an adiabatic path (hence allowing the carbon atom
relaxation to its equilibrium position at every point along the reaction
coordinate) an energy barrier $\sim$0.2 eV high is found, as a consequence
of the re-hybridization. These adsorption curves are shown in fig.\ref{fig:Hpath}.
The barrier (which is also present when the substrate is kept planar) has an important, purely electronic origin. Indeed, 
it has been shown \citep{bonfanti08} that it results from an avoided crossing between a repulsive interaction with the Kekul\'e-like ground-state and an attractive interaction with the low-lying, Dewar-like excited state (see Fig.4 in 
\onlinecite{casolo09}). 
This can be nicely understood in terms of the chemical picture above since the Kekul\'e-like structures do not have unpaired electrons 
which can readily couple with that of the incoming H atom. \\
The overall binding picture of H atoms has found substantial experimental proof for graphite surfaces. Hot hydrogen atoms produced by dissociating $H_2$ molecules at $\sim 2000$ K are required to overcome the barrier and observe sticking. Thus, chemisorption is under \emph{kinetic control}
\footnote{H diffusion is largely impeded by electronic/geometrical effects, see below.}, in marked contrast with vacancy formation through $e^-$/ion bombardment discussed above. Indeed, as we show below, H atoms do \emph{not} adsorb completely random on the surface. A number of TPD, AES, EELS and HREELS spectroscopy data \citep{Zecho2002,Zecho2004,Guttler2004,Guttler2004a,Andree06} is available, along with detailed kinetic Monte Carlo simulations \citep{Gavardi09,Cuppen2008} of TP desorption curves and accurate studies of vibrational relaxation dynamics \citep{sakong} and reaction dynamics to form $H_2$ \citep{Martinazzo2005b,Martinazzo2006a,Martinazzo2006b,Morisset2004,Morisset2005,Jackson&Lemoine2001,lowe09}. 
Notice that even though we focused here on adsorption of H atoms the same holds for other simple, monovalent chemical species. \\

\section{Low density: $\pi$-defect structure}\label{sec:LD}

\subsection{The appearance of midgap states}

The effect of atomic scale defects in graphite, and later on in graphene,
has been experimentally  studied since the late eighties, when scanning 
tunneling microscopy (STM) allowed to capture images 
on solid surfaces at atomic scale resolution. It appeared immediately that when a vacancy was 
created by irradiating the sample, a 
bright $\sqrt{3}$x$\sqrt{3}R30^\circ$ charge density reconstruction appears 
\citep{Mizes1989,Ruffieaux2000,Ugeda10}.\\
A carbon vacancy or a defect in the $\pi$-network due to a monovalent 
chemisorbed species creates in graphene an \emph{imbalance} between the number of sites in
 each sublattices. This lowers the overall lattice symmetry, 
up to eventually remove the Dirac cones and open a band gap.
Looking at the tight-binding Hamiltoninan in equation \ref{eq:TB}
the introduction of a $\pi$-defect in the graphene lattice reads as 
the removal of the basis function corresponding to the defect site,
and the system eigenstates become necessarily odd-numbered. Therefore, 
in the nearest-neighbour approximation, because of the electron-hole symmetry, 
one of the eigenvalues in the energy spectrum necessarily lies at the Fermi level.
This zero-energy state is a singly occupied molecular orbital called \emph{midgap} state, even when a gap is not really present.
When relaxing the nearest-neighbour approximation such state moves from the Fermi level, but remains close to it. 
Its presence is important for the transport properties being responsible for resonant scattering mechanisms.\\
The appearance of midgap states in bipartitic systems has been 
intensively studied in solid state physics because of the implications 
they have for the appearance of magnetism. Inui \emph{et al.} \citep{TheoInui} 
formulated a useful theorem for bipartitic tight-binding models with a sublattice imbalance. 
According to their 
result in any bipartite lattice in which the numbers of 
sublattices sites $n_A$ and $n_B$ are not equal, there are 
at least $\eta=|n_A-n_B|$ linearly 
independent eigenfunctions of the Hamiltonian at zero energy, 
all with null amplitudes on the minority sublattice sites. The proof is simple: for let $N_A>N_B$ and 
$\ket{\psi}=\sum_i \alpha_i\ket{a_i}$ be a trial solution at zero energy. The coefficients $\alpha_i$ need to satisfy $\sum_i\braket{b_j|H|a_i}\alpha_i=0$ for $j=1,..N_B$ which is a set of $N_B$ equations for the $N_A>N_B$ coefficients, with $\eta$ linearly independent solutions. This also shows that $\psi$'s localize on the $A$ lattice sites.
\\
Analogous results have been already known in hydrocarbon chemistry 
for some time. The tight-binding approach described above has been   
 used for decades in quantum chemistry to study 
aromatic hydrocarbons, under the name of H{\"u}ckel method. 
The mathematical properties of the H{\"u}ckel Hamiltonian  
have been formalized in a series of theorems and corollaries in a 
famous book of Dewar \citep{DewarBook}. Bipartite lattices were listed there 
as ``alternant'' hydrocarbons, and the 
emergence of midgap states formally predicted in case of odd-numbered 
alternant hydrocarbons.\\ 
Calculations, both at tight-binding and at higher levels of theory (DFT), 
confirm these expectations: in graphene the zero-energy states originated in this way 
correspond to semilocalized modes around the defect which decay slowly with the distance,
\emph{i.e.} with a  $r^{-1}$ power law \citep{Pereira2006,pereira08a}, a result which has been recently confirmed by experiments \citep{Ugeda10}.
Pereira \emph{et al.} \cite{pereira08a} performed a comprehensive analysis of the effect low-density defects  have on the graphene DOS, 
by using numerical tight-binding calculations for $\sim4\text{x}10^6$ lattice sites and analytic results. 
Analogous results have been found in DFT studies\footnote{The approach used is intrinsically periodic. Therefore, the results are best viewed as referring to defects which are periodically arranged on superlattices with large unit cells.
} 
of isolated vacancies \citep{Yazyev2007} and adatoms \citep{casolo09,Boukhvalov2008}.

\subsection{Chemical resonance formula}
\begin{figure}[t]
  \centering
      \includegraphics[width=1.00\columnwidth]{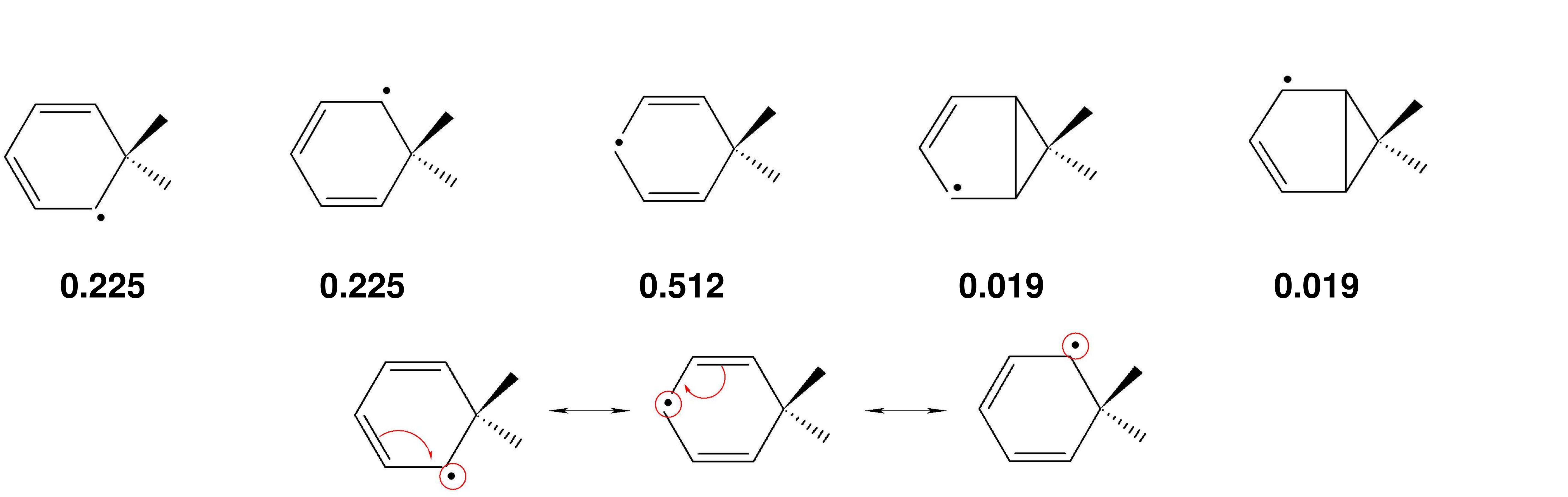}
  \caption{\label{fig:benzmodel}Valence Bond model for the binding of a radical species (a H atom) on benzene.
    The numbers give the weights of the corresponding VB structures \citep{bonfanti08}.  
    The resulting unpaired electron localizes mostly in \emph{ortho} (two leftmost structures on the top row) and \emph{para} (mid panel) position, as emphasized by the bond-switching mechanism reported in the chemical formula of the bottom row. }
\end{figure}
\begin{figure*}[t]	
\centering
\includegraphics[width=0.70\textwidth]{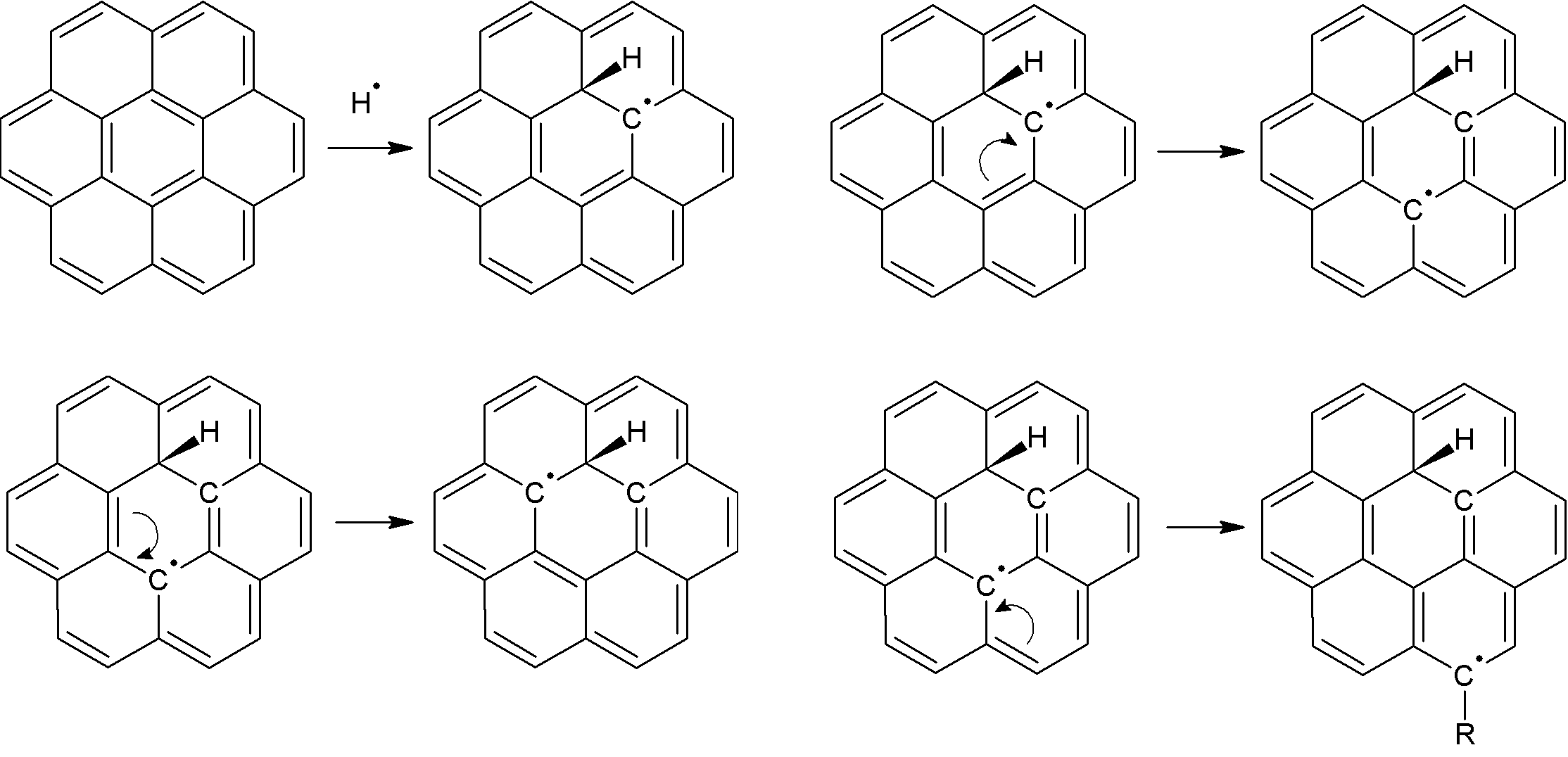}
\caption{Itinerant electron model for the $p_z$-vacancy-induced midgap state}\label{fig:coro}
\end{figure*}
In the case of a single $\pi$-defect, or a random distribution of them, the appearance of midgap states 
may be easily understood by applying the resonance-based VB picture described in Section \ref{sec:2DEG}
. 
Considering benzene as the simplest building block of graphene, it is easy to realize 
how adsorption of a H atom breaks the aromatic network and leaves one unpaired electron 
free to move on the lattice by \emph{bond switching}: 
spin-recoupling with a neighbouring double bond creates an unpaired electron in one every two lattice sites. 
\emph{Ab-initio} VB calculations \citep{bonfanti08} show that this indeed the case: 
the 5 $\pi$ electrons have 5 different ways of couplings (Fig.\ref{fig:benzmodel})
but only those with the unpaired electron in the so-called \emph{ortho} and \emph{para} positions are relevant; an electron in \emph{meta} position would involve a Dewar-like structure, which has a high energy bond-pattern (see Fig.\ref{fig:benzmodel}). 
The bond switching mechanism is very useful and well known in basic organic chemistry, 
where it easily allows predictions for orientation effects, \emph{e.g.} in electrophilic aromatic substitutions. In contrast to the full analysis of possible spin-couplings, 
exporting this model to graphene is then rather straightforward\footnote{This amounts to consider Kekul\'e structures only, which are much fewer than the whole set of $f_S^N$ couplings for all but small $N$ values.}.
A picture of the mechanism is shown in fig.\ref{fig:coro} for a 
coronene model, that is meant to represent the whole graphene lattice. 
The itinerant electron hops between sites of one type only, 
thereby occupying a delocalized state which is the midgap state 
described previously in the tight-binding (MO) picture.\\
Whatever picture we use the result is a \emph{spin density} (magnetization) localized close to the defect, 
on the sites of the hexagonal sublattice not housing it
\footnote{It can also be turned into a \emph{charge-density} by addition/removal of one electron.}. 
At low density, where hybridization does not occur, such spin-density thus determines the appearance of (microscopic) magnetically ordered domains. 
It further influences reactivity of the substrate with foreign species, which can readily `saturate' (singlet-couple) this electron if they land on the correct sites, as will be shown in the next section.
      
Before concluding this Section, we can now understand why simple adatoms do not move on the surface. 
Indeed, for the H atom to hop on the neighbouring site the unpaired spin has to move from one sublattice to the other and this requires breaking completely the existing CH bond and forming a new one: the barrier to diffusion, then, matches the desorption energy. This explains the experimental observation that H atoms are immobile on the surface \citep{Hornekaer2006II}. 
For more complex species, \emph{e.g.} O atoms, spin-recoupling \emph{on} the adatom may help the diffusion (isomerization) process \emph{via} formation of a `bridge` between the two sites. This would explains why DFT computed barriers for diffusion of OH species are definitely smaller than the desorption energy \citep{ghaderi}.  

\section{High density: spin-ordering, clustering  and related issues}\label{sec:HD}

\subsection{Predicting midgap states and magnetism}
\begin{figure*}[t]
  \begin{centering}
    \includegraphics[clip,width=0.85\textwidth]{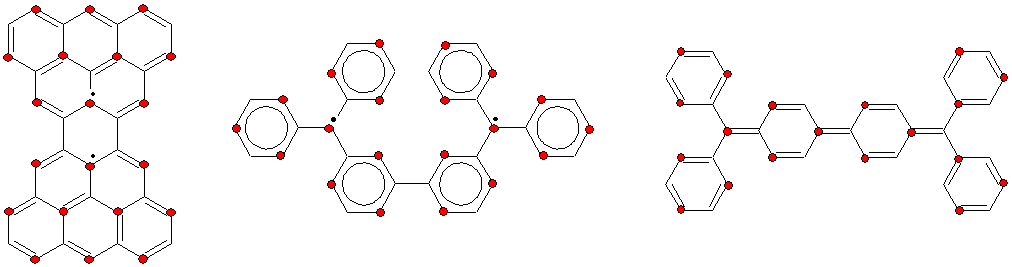} 
    \caption{\label{fig:PAH}  
      Molecules with no sublattice imbalance. The first two molecules on the left are di-radical species, \emph{i.e.} 
      they have two midgap states. The Schlenk-Brauns hydrocarbon shown in the middle panel differs from  
      the Chichibabin hydrocarbon shown in the right panel by the connectivity only. 
      Nevertheless, the latter does not present unpaired electrons. 
      The largest set of non-adjacent sites is marked by red dots.
    }
  \end{centering}
\end{figure*}
The very simple counting rule for midgap states introduced above
usually works fine for graphene,
but fails to be predictive for some class of finite size graphenes 
 or analogous (complementary) holes 
on the graphene sheet. 
For instance, the first two molecules in Fig.\ref{fig:PAH} have no 
sublattice imbalance (being symmetric) but are radical species, 
\emph{i.e.} they necessarily have midgap states \citep{ClarGlobet}.
To a closer inspection, sublattice imbalance is indeed only a sufficient condition 
for midgap states to appear. \\
To rationalize the situation, it is necessary to introduce the concept of 
\emph{non-adjacent} sites in a $N$-site bipartitic system. We say that two sites are non-adjacent  
if they are not bound (connected) to each other; for instance, two sites on the same sublattice are non-adjacent.
Clearly, there exists a maximal set of non-adjacent sites and we call $\alpha$ the sites in this set, and $\beta$ the remaining ones 
($N_{\alpha},N_{\beta}=N-N_{\alpha}$ in number, respectively). 
Each site $\alpha$ binds at least to one site $\beta$, otherwise it would represent a completely isolated site. 
Arranging one electron per site $\alpha$, however, we can form at most $N_{\beta}$ bonds at a time, 
and therefore we are left with $\eta=N_{\alpha}-N_{\beta}=2N_{\alpha}-N$ unpaired electrons, 
\emph{i.e.} midgap states. The case of a sublattice imbalance discussed above is a special result of this rule: 
when $N_A>N_B$, since the A lattice sites are always non-adjacent and $N_{\alpha}\geq N_A$, we have $\eta \geq N_A-N_B$. 
Fig.\ref{fig:PAH} shows molecules having $\eta>0$ and no sublattice imbalance, with the indicated $\alpha$ sets. 
As it is evident from its derivation, this result can be equivalently re-phrased by defining $\eta$ to be the number of unpaired electrons in the Lewis structure(s) with the maximum number of $\pi$ (\emph{i.e.} double) bonds. \\
Notice that, since the spectrum of the Hamiltonian is determined by the system topology the whole set of 
counting rules 
for midgap states can be derived entirely from graph theory.
In particular, midgap states appear as zeros of the characteristic 
polynomial of the adjacency matrix $\bf{A}$, that defines the connectivity 
of the graph \citep{ChemGraphTheo,RandicChemRev}. 
In this context, the above result is known as \emph{graph nullity} theorem. \\
Having derived the exact conditions determining the appearance of midgap states, 
the question arises of how spins couple when a number $\eta$ of unpaired electrons are present. 
The determination of the spin state cannot come, of course, from the simple tight-binding Hamiltonian, 
since in these open-shell configurations energy ordering is mainly determined by electron correlation.
At first glance, it can be guessed that electrons occupying 
quasi-degenerate midgap states tend to keep their spins parallel, 
in a sort of molecular Hund's rule, as this reduces Coulomb repulsion, 
\emph{i.e.} system's total spin should always be\citep{LonguetHiggins} $\eta/2$. 
This is actually the case only when midgap states originate from a sublattice imbalance, 
since in such instance they are forced to stay on the same sublattice. 
When midgap states (unpaired electrons) lie on different sublattice they best couple at \emph{low} spin. 
This result can be shown to be exact for the realistic model provided by the (repulsive) Hubbard Hamiltonian: 
Lieb \cite{TheoLieb} showed that for \emph{any} bipartitic system at half-filling
the ground-state spin $S$ is given by the sublattice imbalance 
$S=\frac{1}{2}|N_A-N_B|$. This is a subtle effect of electron correlation, 
which would lead to an energetically unfavourable spin polarization of the 
remaining occupied orbitals if the above Hund rule were followed in absence of sublattice imbalance 
\footnote{Notice that for two electrons in different zero energy state, first-order perturbation theory always gives a triplet ground-state. The nature of the midgap states and the ensuing interactions with the doubly occupied orbitals play a decisive role.}. 
From a different perspective, it has been associated with the most
``spin-alternant''  structure (Ovchinnikov's rule \cite{Ovchinnikov}).\\ 
According to the rules above it is now possible to predict the 
number of midgap states and the spin state of a number of complex graphene 
structures without relevant exceptions.
We only note that the theorems stated above for bipartite 
lattices do not apply for topological defects that destroy bipartitism. 
Nevertheless, it has been noticed that the Ovchinnikov's rule can be 
usually extended to non-bipartite systems \citep{VBBook}, 
although some care has to be paid \citep{Lopez2009}.
For instance, the ground-state multiplicity of Stone-Wales 
defects is correctly predicted to be zero by this rule.\\

\subsection{Preferential sticking}

When adsorbing hydrogen atoms on graphite or graphene under kinetic control STM images clearly show
the formation of dimers and clusters \citep{Hornekaer2006II}. 
Since H atoms are immobile on the surface this must 
be due to a \emph{preferential sticking} mechanism.
This mechanism was first suggested by 
Hornakaer \emph{et al.}\cite{Hornekaer2006II} who looked at the STM images formed by exposing Highly Oriented Pyrrolitic
Graphite (HOPG) samples to a H atom beam, 
and observed formation of stable pairs, also confirmed by \emph{first-principles}
calculations \citep{Rogeau2006,Hornekaer2006II}.
Later Casolo \emph{et al.}\cite{casolo09} showed that the preference 
for certain lattice sites comes from the spin 
density localized on one of the two sublattices (the midgap state), as
generated by the first adsorbate. \\
The overall picture \citep{casolo09} is consistent with the VB chemical model: when a first H atom 
is on an A-type site, the unpaired electron localizes on the B sublattice and 
bond formation easily occurs on its sites. An ``AB dimer'' (which has no sublattice imbalance) 
is formed and a singlet ground-state is obtained where aromaticity is partially restored. 
Conversely, if adsorbtion occurs on the same sublattice, \emph{i.e.} to form ``A$_2$'' dimers, 
the incoming H atom does \emph{not} make use of the available spin-density, 
and adsorption energies are comparable to that of the first H atom.
Furthermore, as another electron is set free on the \emph{same} B sublattice occupied by the unpaired electron, 
the ground state is a triplet ($\eta=N_B-N_A=2$). \\
\begin{figure*}[t]
  \begin{centering}
  \includegraphics[width=0.7\textwidth]{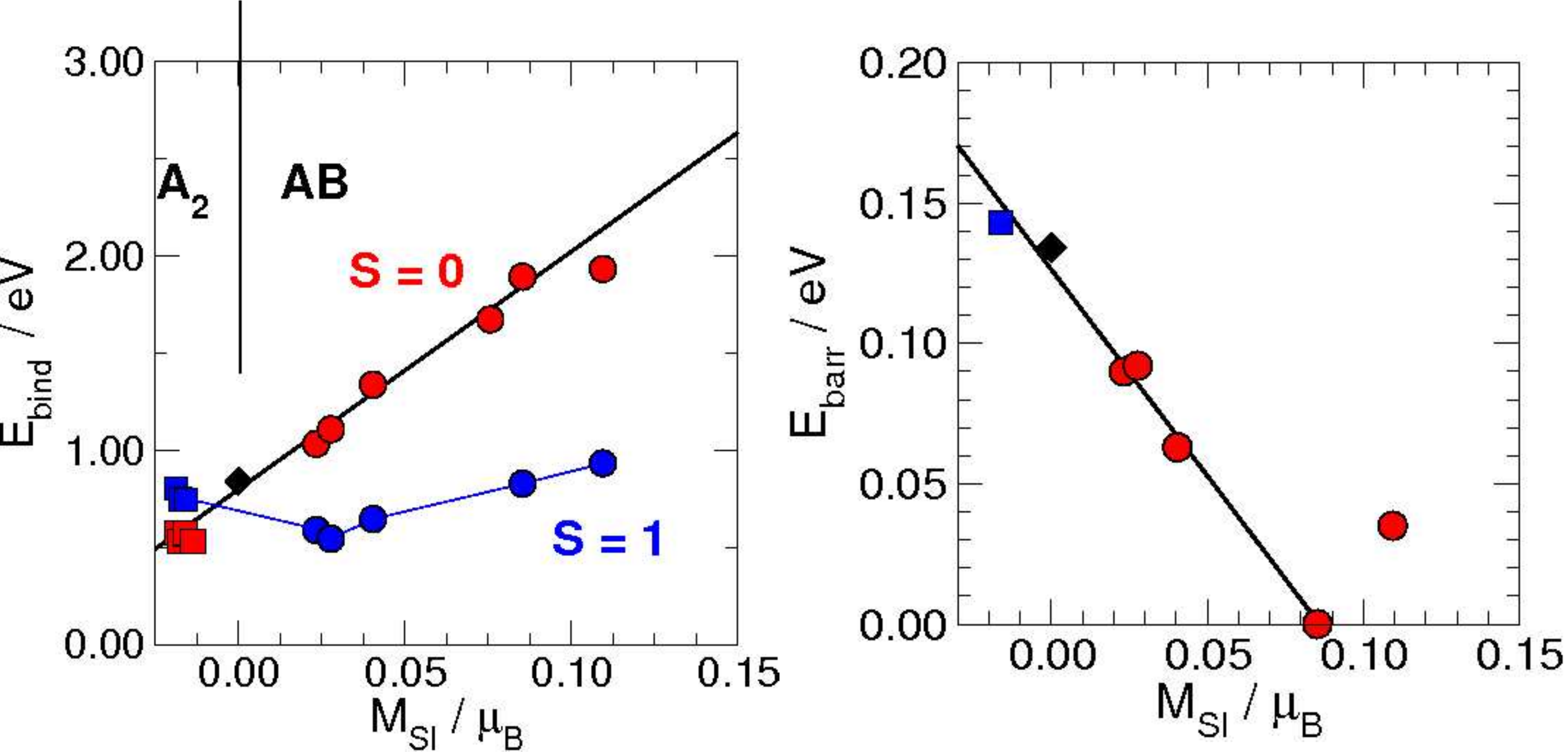}
  \par\end{centering}
  \caption{\label{fig:dimers}
    Left panel: Binding energies for secondary H adsorption
    as a function of the site-integrated magnetization ($M_{SI}$), for $AB$ (
    circles) and $A_2$ (squares) dimers. Both singlet (red) and triplet
    (blue) solutions are shown in red and blue respectively.  
    Also shown the data point for single H adsorption (black diamond) and a linear fit to the data
    set (solid line).
    Right panel: corresponding barrier energies for secondary atom
    adsorption (ground-state only). Data point at M$_{SI}$=0 is for single H adsorption.} 
\end{figure*}
The results of DFT calculations \citep{casolo09} on a number of dimers are shown in Fig.\ref{fig:dimers} 
as function of the \emph{site-integrated} magnetization, 
\emph{i.e.} the average number of unpaired electrons in each site as results from 
the first adsorption event.  
It should be noted that substrate relaxation effects, though substantial ($\sim0.8$ eV), 
are site-independent for all but the \emph{ortho} dimer\footnote{Surface puckering upon adsorption, to a good approximation, involves nearest neighbouring C atoms only.}; thus the curves 
in the graphs of Fig. \ref{fig:dimers} reflects purely electronic effects.
Binding and barrier energies both depend linearly on the local magnetization, 
thereby implying a linear relationship between them; this is a common tendency in
activated chemical reactions known as Br{\o}nsted-Evans-Polayni rule.
An exception is provided by the \emph{ortho} dimer (rightmost data point in the graphs of Fig. \ref{fig:dimers}), 
whose formation requires further rearrangement in the first C-H neighbourhood. 
This is shown in figure \ref{fig:Odimers} where the equilibrium geometry of the dimer is reported in the left panel. 
It is clear from the figure that the two H atoms point in opposite directions (as in a H-C-C-H eclipsed conformation of an alkane), which suggests that, despite their proximity, they would not easily desorb to form $H_2$ upon heating the substrate. 
This is indeed what has been found by a combined theoretical and experimental study by Hornalaer \emph{et al.}\cite{Hornekaer2006}: 
upon heating, the \emph{ortho} dimer prefers to isomerize to the \emph{para} dimer, which dehydrogenates easier (\emph{i.e.} at a lower temperature). The highest temperature peak in the TPD spectra corresponds then to this isomerization process.
The \emph{para} dimer itself, whose equilibrium geometry is shown in the right panel of Fig.\ref{fig:Odimers}, 
forms abundantly when exposing graphene to a H atom beam, since its formation is \emph{barrierless} (see fig.\ref{fig:dimers}). 
This forms the basis for the preferential sticking mechanism first suggested by Horenkaer \emph{et al.}\cite{Hornekaer2006II},
which is here summarized with the results of fig.\ref{fig:dimers}, 
namely formation of $AB$ dimers is both thermodynamically
\emph{and} kinetically favoured over formation of $A_{2}$ dimers
and single atom adsorption.\\       
\begin{figure}[t]
\begin{centering}
\includegraphics[clip,width=0.9\columnwidth]{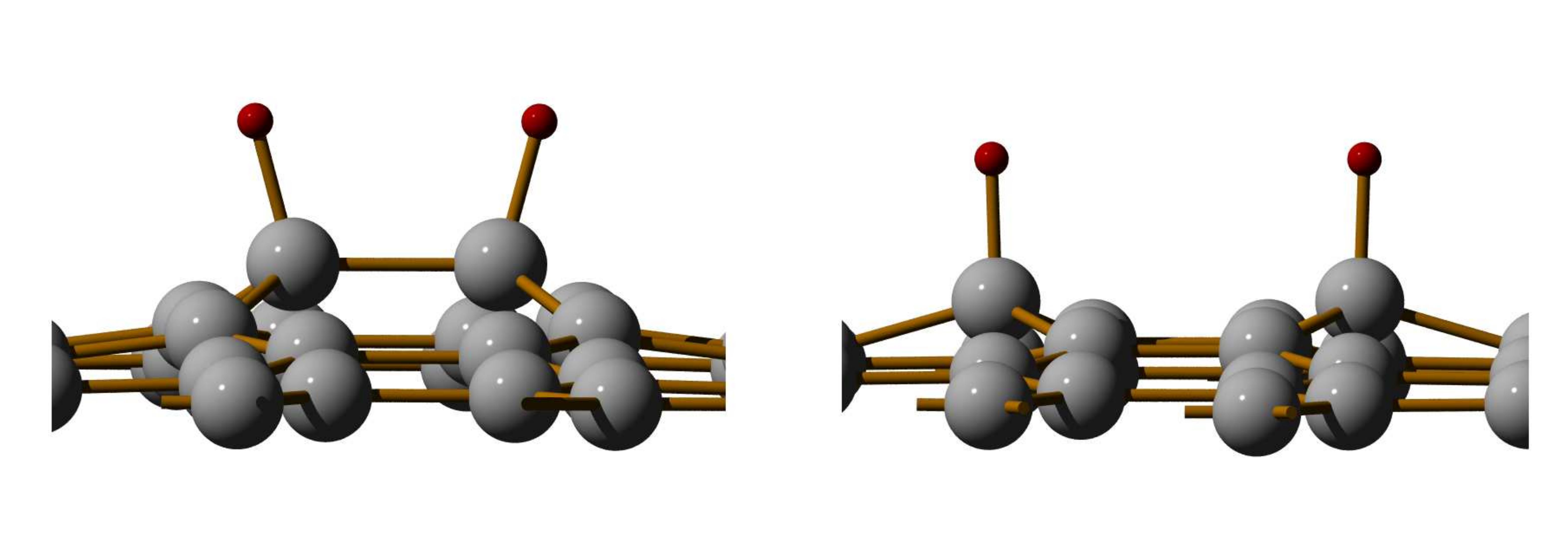} 
\par\end{centering}
\caption{\label{fig:Odimers} Equilibrium geometry for ortho (left) and
  para (right) dimer structures.}
\end{figure}
Unfortunately\footnote{There is a long open search for efficient pathways leading to $H_2$ formation on graphitic surfaces, because of its importance in explaining the observed abundance of molecular hydrogen in the interstellar medium.}, the preferential sticking mechanism above works only for dimers. Once an AB dimer is formed (A$_2$ is unfavoured) there are no further unpaired electrons available, 
and no bias on the adsorption of additional H atoms. This is confirmed 
by DFT calculation \citep{casolo09} on a number of larger A$_{2}$B$_{2}$,
A$_{2}$B, A$_{3}$B$_{1}$ and A$_{3}$ clusters. 
As expected from the VB model above, adsorption of a third hydrogen 
atom to a stable AB dimer parallels that of
the first H, with essentially no preference towards any specific sublattice position, and 
always produce doublet structures ($M=1\,\mu_{B}$) (see fig.\ref{fig:trimers}). 
Similar conclusions hold when adding a 
third H atom to the (magnetic) \emph{meta} dimer A$_2$: adsorption
on B lattice sites is strongly favoured ($E_{bind}=1.2-1.9$ eV)
and produces doublet structures ($M=1\,\mu_{B}$), whereas H atoms
bind to $A$ lattice sites with an energy $\sim0.7-0.8$ eV and produce
highly magnetic structures ($M=3\,\mu_{B}$) (see fig.\ref{fig:trimers}). 
Energy barriers to adsorption follow the same trend: calculations show that, with few
exceptions, barriers to sticking a third H atom compare rather well
with that for single H atom adsorption for the processes AB$\rightarrow$A$_{2}B$
and A$_{2}\rightarrow$A$_{3}$, and may be considerably smaller for
A$_{2}\rightarrow$A$_{2}$B ones. These three-atom clusters, 
similarly to the single H atom, necessarily bias the adsorption 
of a fourth atom. The computed binding and barrier energies for this process have been found to compare
rather well with the dimer values, actually they nicely fit to the \emph{same} linear
trends shown in Fig.\ref{fig:dimers}.
Finally, all the considered   A$_{2}$B$_{2}$,
A$_{2}$B, A$_{3}$B$_{1}$ and A$_{3}$ clusters have been found to have 
0, 1, 2 and 3 unpaired electrons in their ground-state, respectively,
in agreement with expectations (\emph{i.e.} either the VB model or the Lieb theorem). \\
Few exceptions to this picture are for compact clusters where
substrate relaxation does play some role, see \emph{e.g.} the structures on the right of fig. \ref{fig:trimers}. 
Compared to other trimers, these structures are favoured because of the substrate 
\emph{softening} occurring after formation of the \emph{para} dimer, 
which is in a typical \emph{boat} configuration (fig.\ref{fig:Odimers}).
Such softening is expected to reduce the relaxation energy needed for the binding of the additional atom, 
with a gain of some tenths of eV on the overall energetic balance\footnote{Remember that the relaxation energy for the single H atom ($\sim0.8$ eV) has the same magnitude as the overall binding energy, \emph{i.e.} about half of the bond formation energy is spent for relaxing the substrate.}. 
This would explain why experiments at intense H atom flux  
do not find a random distribution of \emph{dimers}, as would be expected on basis of electronic effects only, 
rather clusters made up of a number of atoms \citep{Hornekaer2007,Ferro09}.
Though a detailed analysis would require the knowledge of the adsorption \emph{barrier} for a rather large number 
of clusters, the linear relationship shown above may help in 
making educated guesses on the basis of the binding energies only.\\
\begin{figure*}[t]
\begin{centering}
  \includegraphics[width=0.98\textwidth]{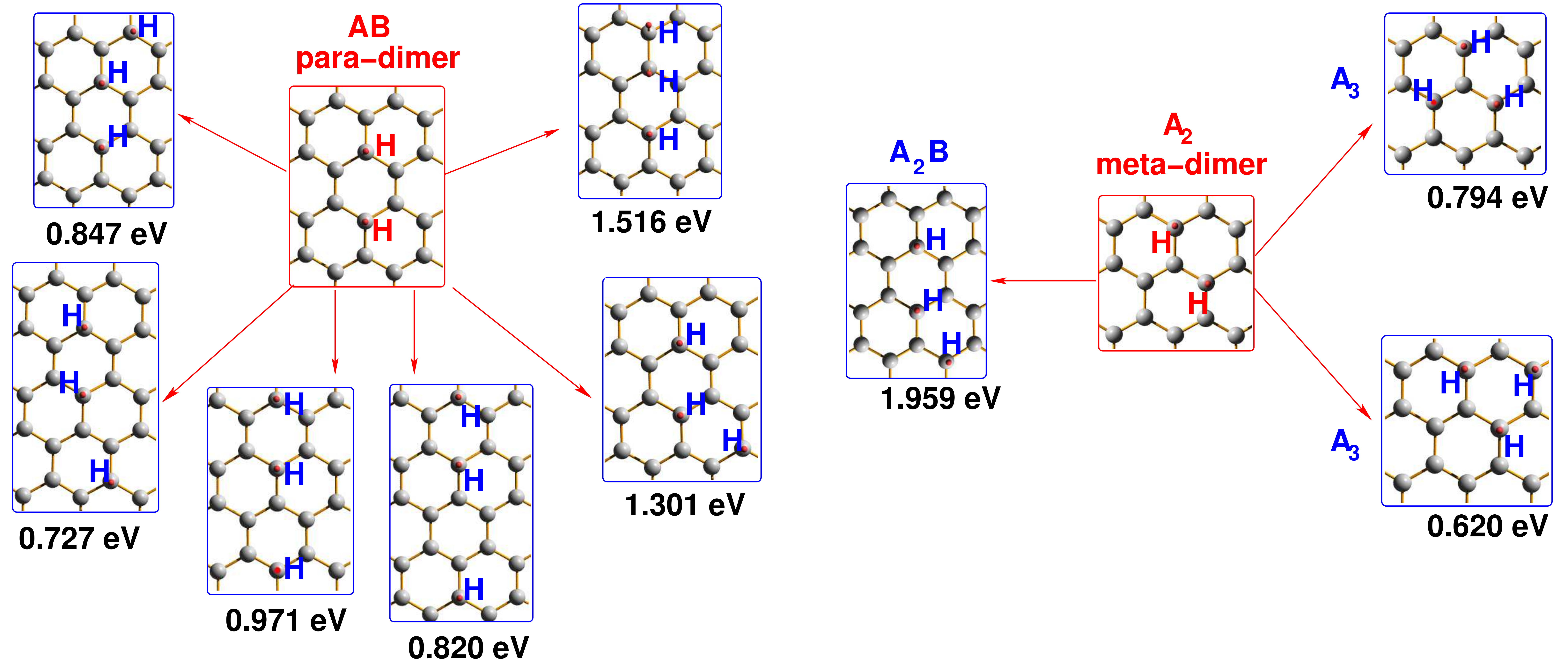}
\par\end{centering}
\caption{\label{fig:trimers} Some of the possible hydrogen trimers
  structures obtainable from the AB \emph{para} dimer (left diagram) 
  and from the A$_2$ \emph{meta} dimer (right). Binding energies are also shown. }  
\end{figure*}

\subsection{A route to graphane?}

\emph{Graphane} is a novel two-dimensional material, namely the
fully hydrogenated graphene sheet. 
The structure is still bipartitic and each sublattice 
bears all hydrogens on the \emph{same} side of the lattice plane (in \emph{meta} to each other),
in such a way to have a chair-like configuration as in cyclohexane. It is an insulating material with
no $\pi$ electrons \citep{sofo07,Lebegue09}, that might have been recently obtained 
by simply exposing graphene to cold hydrogen plasma \citep{Elias09}. 
This result is very interesting in light of what has been shown in 
the previous sections. In order to produce graphane by simple 
hydrogen exposure it is necessary that either H atoms adsorb 
selectively on one sublattice \emph{only} for a given graphene face
or hydrogen diffuses (isomerize) to occupy the sites on the right \emph{face}.\\
According to the discussion of previous section, formation of \emph{meta} dimers
is unlikely, \emph{ortho}, \emph{para} positions being highly favoured for adsorption.  
According to DFT calculations on free-standing graphene \citep{casoloTesi}, 
this is true both for the \emph{syn}- (on the same face) 
and the \emph{anti}- (on opposite faces) dimers. 
However, even if hydrogen atoms were likely to form \emph{anti-para} dimers also in supported graphene 
(and they are 
required for graphane production) an efficient \emph{syn-} to \emph{anti-} conversion mechanism would be needed to convert
those dimers already formed on the same graphene face.
Unfortunately, as we have seen in the previous section, even in this fortunate case, 
no true preferential sticking can occur after dimer formation and it will be very unlikely that all the other
hydrogens will chemisorbed in the correct sites and face. 
Indeed, recent molecular dynamics simulations showed that disordered, 
frustrated hydrogenated domains would rather form \citep{flores09}. \\
If graphane has been really formed by hydrogen exposure some other effect has to
play a role. Curved graphene areas might help this process. Graphene is a very
elastic membrane that naturally exhibit ripples that tend 
to lay down along steps and kinks of the supporting substrate on which it was
grown. Indeed recent experimental findings suggested that hydrogen 
chemisorbs more efficiently on the ridges of the silicon carbide substrate
surface onto which graphene usually lays \citep{Hornekaer2009}. 
Moreover, it has been also shown that hydrogenation of single-layer graphene is
easier than for many-layer graphene, likely as a consequence of the higher 
corrugation displayed by the graphene surface \citep{Luo09,Brus09}. 
This is reasonable, as for nanotubes the curvature reduces the
$p_z$-$p_z$  overlap, \emph{i.e.} aromaticity, thereby lowering the barrier 
energies for H chemisorption \citep{Ruffieux2002,Ruffieux02}. 
Still, there is no clear evidence that local curvature plays a role in graphane formation, 
and more investigations in this direction are needed.\\

\section{Defect-based material design}\label{sec:superlattices}

When it comes to device fabrication only few of the many extraordinary 
properties of graphene are relevant, at least for the chip-makers \citep{Schwierz2010}. 
Among them, its thickness allows the thinnest possible gate-controlled regions in transistors 
and, according to scaling theories, should reduce electrostatic problems if short channels
have to be built. Mobility is an important factor as it allows for instance high-performance interconnects and
fast response to external (gate) potentials. It becomes of secondary importance  
in short channels, where high fields build up and carrier velocity saturates, but also in this respect 
graphene proved to have superior properties than conventional materials. 
Indeed, high-performance transistors for frequency applications have been
realized \citep{avouris10}, and record cut-off frequencies are being continuously scored. 
However, for its usage in logic applications the absence of a band-gap 
is a major problem \citep{Schwierz2010,Avouris07}: 
even when the Fermi level crosses the charge neutrality point a
non-zero residual conductivity avoids the complete current
pinch-off \footnote{The defects discussed in this chapter, along with charged scatterer, 
are ascending as the most likely origin of the conductivity minimum \citep{peres10}. The counter-intuitive role of defects in \emph{increasing} the conductivity finds its origin in the modification of the graphene DOS close to the Dirac point. }. 
The absence of a band-gap, indeed, prevents the achievement of the high
current on-off ratios required for logic operations. \\
Graphene can be turned into a true semiconductor by properly engineering it. 
Electron confinement, though in general not trivial for massless, pseudorelativistic carriers, 
can be obtained by cutting large-area graphene to form narrow nanoribbons. 
Apart from related fabrication issues, one main drawback of such an approach is the removal 
of the Dirac cones and the resulting band-bending. This is expected to increase the effective mass of the carriers,  
thereby reducing their mobility. Indeed, it has been generally found that mobility 
is a decreasing function of the gap \citep{Schwierz2010}, and this is an undesirable by-side effect 
worth considering with such a traditional approach.    
Alternatively, \emph{symmetry breaking} is known to turn the massless Dirac carriers into massive (yet pseudorelativistic) carriers.
This can be realized by depositing or growing graphene on a substrate that renders inequivalent the two sublattice positions. 
For instance, boron nitride has the same honeycomb lattice
as graphene and a similar cell parameter, but presents
two inequivalent sublattices. When graphene is in contact
with such a surface B and N interact differently with
the carbon atoms of the graphene sheet, breaking its sublattice
equivalence and lifting the degeneracy of the two
bands. A similar situation is achieved for graphene grown on silicon carbide surfaces, 
where a gap has been observed by angle-resolved photoemission spectroscopy 
though subtle electron correlation effects may play a role in such case \citep{zhou07,Bostwick07}.\\
In the following sections we describe alternative possibilities for opening a gap in graphene band structure, 
namely those offered by superlattices of defects and dopants. One interesting finding in this context is the
proof that a band-gap can be opened in graphene \emph{without} breaking its symmetry,  
with the advantage the new Dirac cones (massless carriers) appear right close 
to the gapped region \citep{martinazzo10}. For this reason we start introducing some symmetry considerations, 
extending the arguments given in section \ref{sec:2DEG}. 

\subsection{Symmetry considerations}

\begin{figure*}[t]
\begin{centering}
  \includegraphics[width=0.7\textwidth]{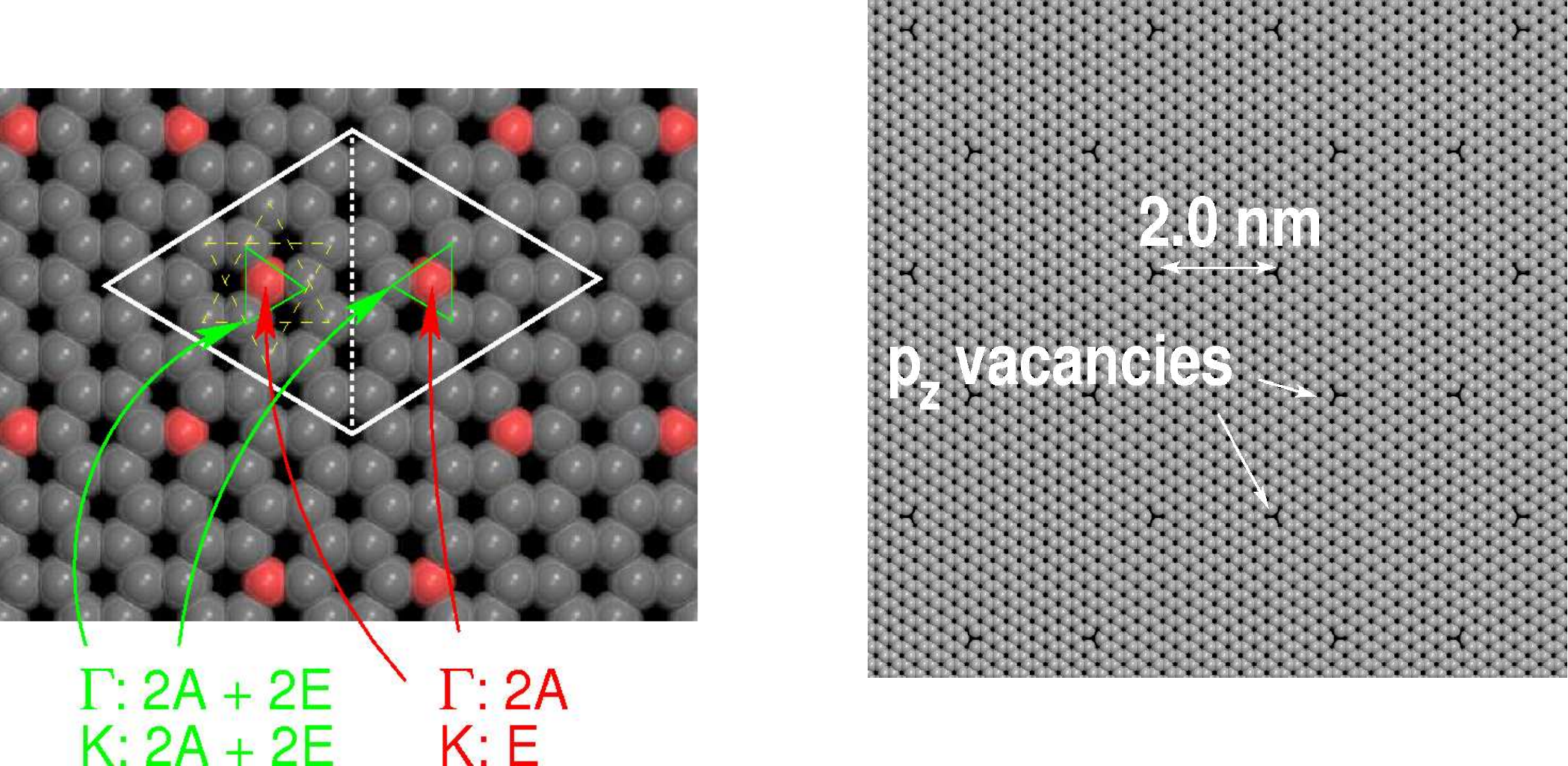} 
\par\end{centering}
\caption{\label{fig:honeycombs} Left: counting the number of irreps generated by the atomic 
basis in a $4\text x4$ supercell: indicated are the irreps generated by the atoms at the center of the half-cells (red balls)
and by green triangles. Right: a simple ``supergraphene'', the simplest defective $14\text x 14$ honeycomb.}
\end{figure*}

As we have seen in section \ref{sec:2DEG} graphene's unconventional electronic 
properties are strictly related to its $D_{6h}$ point symmetry. The \emph{k}-group 
at the K-K' high-symmetry points ($D_{3h}$) allows for doubly degenerate
irreducible representations, and Bloch functions built
with $p_z$ orbitals of A and B sublattices span just one of its two-dimensional irreps.
As $e-h$ symmetry does not mix one- ($A$) and two- ($E$) dimensional irreps this level has to lie
at zero energy, where the Fermi level ($\epsilon_F$) is located.\\ 
Were not there such degenerate level, graphene would be, 
as any other bipartitic system at half-filling, semiconducting.  
Graphene can be forced to be so by either lowering the symmetry 
(\emph{i.e.} changing the \emph{k group} at K(K') to a simpler one), 
or changing the \emph{number} of $E$ irreps at the special points while keeping the overall symmetry.  
In the latter, more intriguing case, since the overall point symmetry is preserved, 
degeneracies may still occur at energies different from $\epsilon_F$, 
and new Dirac cones are to be expected.  
The ``recipe'' for doing that is very simple \citep{martinazzo10}: 
$n\text x n$ graphene superlattices have the same symmetry properties  
and $2n^2$ atoms per cell; by symmetrical removing a number of C atoms
\footnote{As shown in the previous sections one can equivalently 
introduce either a vacancy or an adatom.} is possible
to change the number of irreps and turn, in particular, the $E$ ones to be \emph{even} at every, 
highly symmetric special point ($\Gamma,K,K'$). With few exceptions of residual accidental degeneracies, 
this opens a gap in the band structure. \\
The approach is made effective by counting the number of irreps generated by $2n^2$ atoms in a $n\text{x}n$ unit supercell. 
The results of this calculation \citep{martinazzo10} 
can be grouped
into three different sequences, $n=3m,3m+1,3m+2$ ($m$ integer), 
according to the BZ folding properties. 
In two thirds of the cases, \emph{i.e.} when 
$n=3m+1,3m+2$
, removal of the atoms at 
the center of the two-half cells (red balls in the left panel of fig.\ref{fig:honeycombs}) 
is sufficient for opening a gap. 
Figure \ref{fig:honeycombs} (right panel) shows one of the simplest resulting semiconducting superlattice, 
namely with simple, atomic-scale defects arranged in a \emph{honeycomb} lattice. 
Because of their nature, they are best considered as \emph{supergraphenes}.

\subsection{Superlattices of vacancies or holes}

\begin{figure*}[t]
\begin{centering}
  \includegraphics[width=0.3\textwidth]{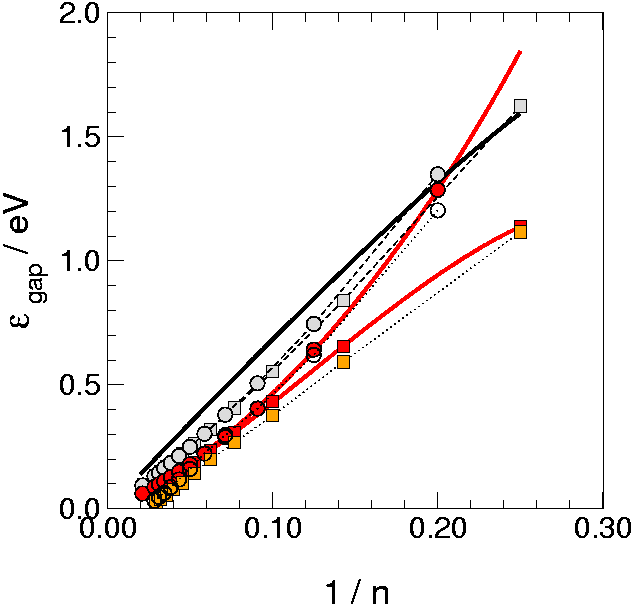}
  \hspace{0.1cm}
  \includegraphics[width=0.3\textwidth]{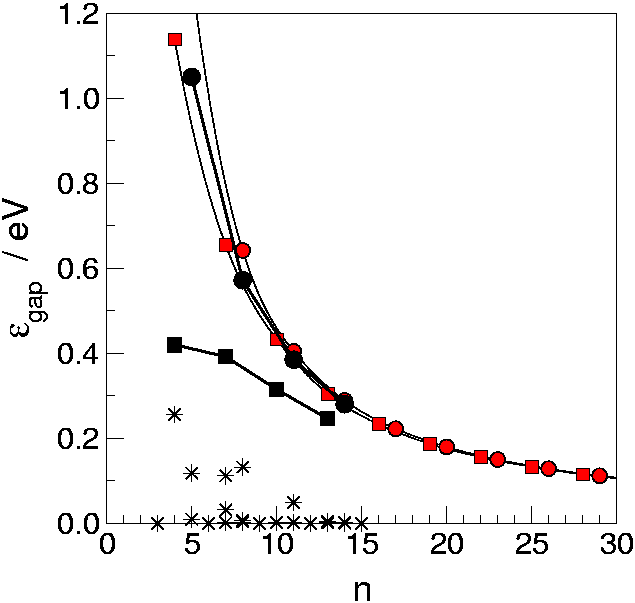}
  \hspace{0.1cm}
  \includegraphics[width=0.3\textwidth]{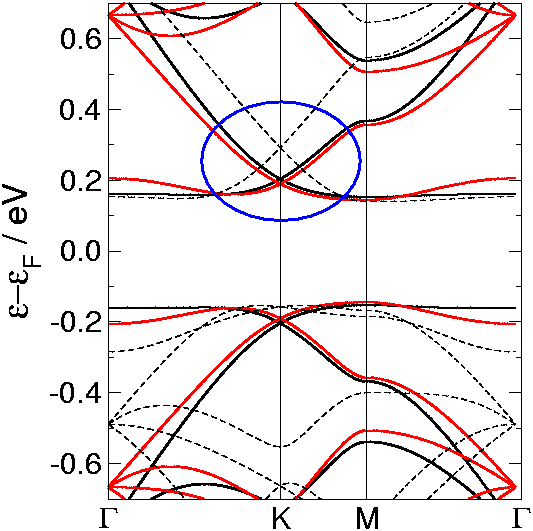}
\par\end{centering}
\caption{\label{fig:gaps} Energy gaps in simple supergraphenes made with $p_z$ vacancies. Left: results of TB calculations as functions of $1/n$. The symbols are for different parameters of the TB hamiltonian, and the solid line is the result of a perturbative calculation at the K point. See \cite{martinazzo10} for details. Middle: tight-binding (red) \emph{vs.} DFT (black) results. In the latter case, defects have been modelled as H atoms. Stars represent the results for asymmetric dimers placed in the same $n\text xn$ supercells. Right: energy bands for the $n=13,14$ supergraphenes.}
\end{figure*}

Tight-binding and DFT calculations on the simple structures identified in the previous section 
show indeed a sizable band-gap. The gap size approximately scales as $v_{F}/l_{n}$ where $v_{F}$ is the Fermi velocity in pristine
graphene and $l_{n}$ is the distance between defects ($l_{n}=na/\sqrt{3}$), 
as can be guessed from a dimensional analysis or obtained from 
a perturbative calculation
within the tight-binding approach \citep{martinazzo10}. Both the size and 
the scaling compare favourably with the gap in armchair nanoribbons \citep{louie06}.
However, 
one distinctive feature of such structures is the additional presence of new Dirac cones right
close to the gapped region (blue circle in fig.\ref{fig:gaps}). This might be important in charge transport, 
since they can sustain massless carriers when the Fermi level, as tuned by a gate potential, 
is swept across the gap. 

\begin{figure}[t]
\begin{centering}
    \includegraphics[width=0.9\columnwidth]{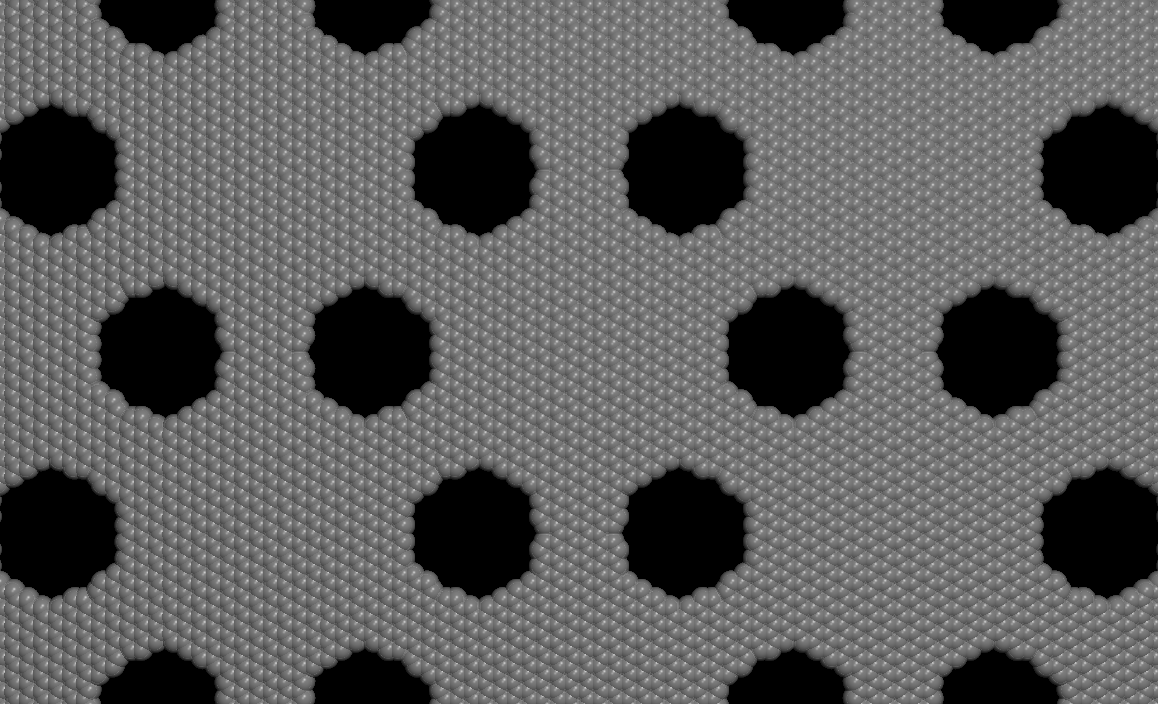}
  \par\end{centering}
\caption{\label{fig:antidot} A honeycomb antidot superlattice presenting a gap by the symmetry preserving approach discussed in the text.}
\end{figure}

In practice, it is  still experimentally challenging 
to realize the atomic-scale patterned structures introduced above. 
It is however sufficient to consider similar superlattices of \emph{holes} analogously 
to the graphene \emph{antidots} superlattices \cite{Antidot1,Antidot2,Antidot3}.
The resulting structures are \emph{honeycombs antidots} as the one shown in fig.\ref{fig:antidot}.    
They are experimentally feasible, 
since it has been shown possible \cite{meyer08,fischbein08,antidot08,weiss09,GrapheneHoles} 
to create circular holes with diameters as small as $2-3$ nm and periodicity
$\sim5$ nm. Analogous patterns of H adatoms have also been realized  
thanks to the interaction between graphene and an underlying metal surface  
that creates Moire patterns activating chemisorption in specific areas
\citep{HornekaerNature}.\\
Tight-binding calculations on the honeycomb antidots show that the band-gap
is quite large for reasonable values of the superlattice constant and of the hole diameter and,
as before, new Dirac cones appear at low energies, close to the gapped region \citep{martinazzo10}. 
The gap size is even larger than for the simplest structures considered above, 
though the latter remain \emph{optimal} in this context: 
when the gap size is renormalized to the number of defects per unit cell, 
honeycombs such as that reported in fig. \ref{fig:honeycombs} prove to be ``magic''.\\
Notice that previously suggested antidot superlattices \citep{Antidot1,Antidot2}
show comparable gaps, and are therefore equally valid 
candidates for turning graphene into a true semiconductor.
The only difference is a subtle symmetry-related issue. 
These hexagonal superlattices are all of 
$\sqrt{3}n\text x\sqrt{3}n$ type, hence with the same $D_{6h}$ symmetry of the honeycomb lattices considered here, 
and this would suggest that exactly the same results hold for them. 
A closer inspection, however, reveals that in the $\sqrt{3}n\text x\sqrt{3}n$ case, 
the K, K' points of pristine graphene always fold to $\Gamma$. This is advantageous for the band-gap opening, 
since these structures are generally semiconducting if sufficiently defective\footnote{
The number of $E$ irreps is always even because of the `coalescence' of the two valleys. 
Thus, one only needs to remove the accidental degeneracy created by such folding.}. 
However, as in $\Gamma$ the \emph{k} group has inversion symmetry, 
residual degeneracies at $\epsilon \neq \epsilon_F$, is \emph{not} lifted at first order, 
and therefore no linear dispersion is present. 

\subsection{Superlattices of substitutional atoms}

Symmetry arguments similar to the one given above apply as well to the case where
C atoms are replaced (rather than removed) by other species, in such a way to form superlattices
substitutional dopants. The only difference is that now foreign species are present 
and point symmetry can be altered. Here we focus on group IIIA and VA elements, 
mainly because of the fast progresses in methods for the controlled synthesis of N- and B- doped graphenes.
For instance, Panchakarla et al \cite{Panchakarla09} have recently shown how it is possible 
to insert B or N dopants in graphene by adding the correct precursors in the arc discharge chamber, 
while Ci et al. \cite{ciNature10} have reported the synthesis of large islands of boron nitride 
embedded in graphene by atomic layer deposition techniques. Methods to selectively replace C atoms from graphene lattice have also been proposed by 
Pontes et al\cite{Pontes09}.\\
Substitutional defects behave similarly to $p_z$ vacancies 
(to which they reduce when the hoppings become zero) but introduce 
impurity bands which partially hybridize with those of the substrate. 
In addition, the diagonal disorder they introduce breaks $e-h$ symmetry 
giving rise to a Fermi level shift, \emph{i.e.} to $p-$ and $n-$ doping 
for group IIIA and VA elements, respectively. 
If superstructures are only weakly defective, however, the Fermi level shift scales
as $1/n$, since the linear-energy dispersion implies $E_F=v_F\sqrt{\pi n_e }$ 
(here $n_e$ is the electron (hole) excess density, $n_e\propto1/n^2$). 
Thus, analogously to the superlattices of the previous section, 
the defect-induced perturbation affects the electronic structure \emph{close} to the Fermi level, 
and symmetry arguments may be used to establish whether degeneracy occurs at the special points 
in the important low-energy region.\\
It has been shown that, depending on the overall symmetry, 
superlattices of N and B defects (and mixture thereof) can either preserve the Dirac cones
($D_{6h}$ superlattices) or open a band gap ($D_{3h}$) \citep{casolo10II}.
For instance, honeycomb superlattices of B (or N) dopants only ($D_{6h}$ symmetry)
are found to preserve the Dirac cones, and to be only $p-$ ($n-$) doped 
(the cone-apex shift being $\propto 1/n$). 
Indeed, the low-energy band structure in these superlattice is very similar to that of graphene, 
with a group velocity at the shifted Dirac cones depending linearly on $1/n$ too. 
As the Fermi level can be tuned by a gate potential, these systems offer the possibility
of investigating the role that the effective speed of light (the above velocity at the Dirac cones)   
has on the transport and optical properties of graphene. 
Conversely, BN-honeycomb superlattices ($D_{3h}$ symmetry), thanks to the balanced doping,
are found to develop a gap right at the Fermi level. The resulting dispersion relation is found to be quasi-conical,
corresponding to \emph{massive} Dirac fermions. The resulting gaps are found sizable and, furthermore,
the effective rest masses are rather small, $m^*\sim 0.01 m_e$ or smaller. 
This suggests that these structures might have good field-switching and transport properties.

\bibliographystyle{jcp}

\begin{thebibliography}{100}

\bibitem{wallace47}
{\sc P.~R. Wallace},
\newblock {\em Phys. Rev.} {\bf 71}, 622 (1947).

\bibitem{castroneto09}
{\sc A.~H. {Castro Neto}}, {\sc F.~Guinea}, {\sc N.~M.~R. Peres}, {\sc K.~S.
  Novoselov}, and {\sc A.~K. Geim},
\newblock {\em Rev. Mod. Phys.} {\bf 81}, 109 (2009).

\bibitem{Bena09}
{\sc C.~Bena} and {\sc G.~Montambaux},
\newblock {\em New J. Phys.} {\bf 11}, 095003 (2009).

\bibitem{peres10}
{\sc N.~M.~R. Peres},
\newblock {\em Rev. Mod. Phys.} {\bf 82}, 2673 (2010).

\bibitem{Note1}
Interestingly, this operation corresponds to an operator $\protect \mathaccentV
  {hat}05E{\pi }$, $\protect \mathaccentV {hat}05E{\pi } \protect \mathaccentV
  {hat}05E{c}_{i,\sigma }=(-)^\tau \protect \mathaccentV {hat}05E{c}_{i,\sigma
  }\protect \mathaccentV {hat}05E{\pi }$ ($\tau =1,2$ for $c=a,b$), which
  reduces to the $z$ component of the pseudospin in spinor notation.

\bibitem{mirman}
{\sc R.~Mirman},
\newblock {\em Point groups, space groups, crystals, molecules},
\newblock World Scientific, Ltd, Singapore, 1999.

\bibitem{Note2}
The remaining symmetry elements determine the so-called \protect \emph {star}
  of the given $k$ point, which is the set of points generated by these
  elements once applied to ${\protect \bf k}$. Such set of physically distinct
  points in $k$ space are degenerate in energy; this is the case of $K'$ and
  $K$, for instance, since they belong to the star of each other.

\bibitem{Note3}
In the chemical community is dubbed Parisier-Parr-Pople approximation, after
  Pariser-Parr and Pople who first introduced it in the early fifties.

\bibitem{Note4}
This is so because even the simplest VB wavefunctions can be re-written as
  linear combinations of Slater determinants, and include the so-called
  `static' correlation. The latter is essential for describing bond formation
  and near-degeneracies; in extended systems is responsible for Mott
  transitions.

\bibitem{tantardini85}
{\sc M.~Raimondi}, {\sc M.~Simonetta}, and {\sc G.~F. Tantardini},
\newblock {\em Comp. Phys. Rep.} {\bf 2}, 171 (1985).

\bibitem{Cooper87}
{\sc D.~L. Cooper}, {\sc J.~Gerratt}, and {\sc M.~Raimondi},
\newblock Modern valence bond theory,
\newblock in {\em Ab initio methods in quantum chemistry II}, edited by {\sc
  K.~P. Lawley}, John Wiley \& Sons Ltd., 1987.

\bibitem{cooperbook02}
{\sc D.~L. Cooper},
\newblock {\em Valence Bond Theory - Theoretical and computational chemistry
  10},
\newblock Elsevier, 2002.

\bibitem{VBBook}
{\sc S.~Shaik} and {\sc P.~C. Hiberty},
\newblock {\em A chemist's guide to valence bond theory},
\newblock Wiley, 2007.

\bibitem{Note5}
The number of linearly independent spin-function for $N$ electrons in the $S$
  spin state, usually denoted as $f_S^N$, can be easily obtained by angular
  momentum coupling rules. The properties of the corresponding spin spaces stem
  from their deep connection to the group of permutations of $N$ objects.

\bibitem{Note6}
It is enough to consider the so-called Coulson-Fisher wavefunction.

\bibitem{clar10}
{\sc T.~Wassmann}, {\sc A.~P. Seitsonen}, {\sc A.~M. Saitta}, {\sc M.~Lazzeri},
  and {\sc F.~Mauri},
\newblock {\em J. Am. Chem. Soc.} {\bf {132}}, 3440 ({2010}).

\bibitem{Note7}
With this we mean point defects, \protect \emph {i.e.} adatoms, substituted or
  displaced atoms. It should be noticed, however, that other defects may also
  be important for the transport properties.

\bibitem{KrasheninnikovBigRev}
{\sc A.~V. Krasheninnikov} and {\sc K.~Nordlund},
\newblock {\em J. Appl. Phys.} {\bf 107}, 071301 (2010).

\bibitem{Zettl08}
{\sc J.~C. Meyer}, {\sc C.~O. Girit}, {\sc M.~F. Crommie}, and {\sc A.~Zettl},
\newblock {\em Appl. Phys. Lett.} {\bf 92}, 123110 (2008).

\bibitem{Meyer2010}
{\sc J.~C. Meyer}, {\sc C.~Kisielowski}, {\sc R.~Erni}, {\sc M.~D. Rossell},
  {\sc M.~F. Crommie}, , and {\sc A.~Zettl},
\newblock {\em Nano Lett.} {\bf 8}, 3582 (2010).

\bibitem{ElBarbary03}
{\sc A.~A. {El-Barbary}}, {\sc R.~H. Telling}, {\sc C.~P. Ewels}, and {\sc
  M.~I.~H. nd~P~R~Briddon},
\newblock {\em Phys. Rev. B} {\bf 68}, 144107 (2003).

\bibitem{Yazyev2007}
{\sc O.~V. Yazyev} and {\sc L.~Helm},
\newblock {\em Phys. Rev. B} {\bf 75}, 125408 (2007).

\bibitem{Lethinen2004}
{\sc P.~O. Lehtinen}, {\sc A.~S. Foster}, {\sc Y.~Ma}, {\sc A.~V.
  Krasheninnikov}, and {\sc R.~M. Nieminen},
\newblock {\em Phys. Rev. Lett.} {\bf 93}, 187202 (2004).

\bibitem{Ghio1980}
{\sc E.~Ghio}, {\sc L.~Mattera}, {\sc C.~Salvo}, {\sc F.~Tommasini}, and {\sc
  U.~Valbusa},
\newblock {\em J. Chem. Phys.} {\bf 73}, 556 (1980).

\bibitem{Bonfanti2006}
{\sc M.~Bonfanti}, {\sc R.~Martinazzo}, {\sc G.~F. Tantardini}, and {\sc
  A.~Ponti},
\newblock {\em J. Phys. Chem. C} {\bf 111}, 5825 (2007).

\bibitem{Jeloaica1999}
{\sc L.~Jeloaica} and {\sc V.~Sidis},
\newblock {\em Chem. Phys. Lett.} {\bf 300}, 157 (1999).

\bibitem{Sha&Jackson2002}
{\sc X.~Sha} and {\sc B.~Jackson},
\newblock {\em Surf. Sci.} {\bf 496}, 318 (2002).

\bibitem{casoloSIE}
{\sc S.~Casolo}, {\sc E.~{Flage-Larsen}}, {\sc O.~M. L{\o}vvik}, {\sc G.~R.
  Darling}, and {\sc G.~F. Tantardini},
\newblock {\em Phys. Rev. B} {\bf 81}, 205412 (2010).

\bibitem{bonfanti08}
{\sc M.~Bonfanti}, {\sc S.~Casolo}, and {\sc R.~Martinazzo},
\newblock Unpublished, 2008.

\bibitem{casolo09}
{\sc S.~Casolo}, {\sc O.~M. L{\o}vvik}, {\sc R.~Martinazzo}, and {\sc G.~F.
  Tantardini},
\newblock {\em J. Chem. Phys.} {\bf 130}, 054704 (2009).

\bibitem{Note8}
H diffusion is largely impeded by electronic/geometrical effects, see below.

\bibitem{Zecho2002}
{\sc T.~Zecho}, {\sc A.~G{\"u}ttler}, {\sc X.~Sha}, {\sc B.~Jackson}, and {\sc
  J.~K{\"u}ppers},
\newblock {\em J. Chem. Phys.} {\bf 117}, 8486 (2002).

\bibitem{Zecho2004}
{\sc T.~Zecho}, {\sc A.~G{\"u}ttler}, and {\sc J.~K{\"u}ppers},
\newblock {\em Carbon} {\bf 42}, 609 (2004).

\bibitem{Guttler2004}
{\sc A.~G\"uttler}, {\sc T.~Zecho}, and {\sc J.~K{\"u}ppers},
\newblock {\em Surf. Sci.} {\bf 570}, 218 (2004).

\bibitem{Guttler2004a}
{\sc A.~G\"uttler}, {\sc T.~Zecho}, and {\sc J.~K{\"u}ppers},
\newblock {\em Chem. Phys. Lett.} {\bf 395}, 171 (2004).

\bibitem{Andree06}
{\sc A.~Andree}, {\sc M.~{Le Lay}}, {\sc T.~Zecho}, and {\sc J.~K{\"u}ppers},
\newblock {\em Chem. Phys. Lett.} {\bf 425}, 99 (2006).

\bibitem{Gavardi09}
{\sc E.~Gavardi}, {\sc H.~M. Cuppen}, and {\sc L.~Hornek{\ae}r},
\newblock {\em Chem. Phys. Lett.} {\bf 477}, 285 (2009).

\bibitem{Cuppen2008}
{\sc H.~M. Cuppen} and {\sc L.~Hornek{\ae}r},
\newblock {\em J. Chem. Phys.} {\bf 128}, 174707 (2008).

\bibitem{sakong}
{\sc S.~Sakong} and {\sc P.~Kratzer},
\newblock {\em J. Chem. Phys.} {\bf 133}, 054505 (2010).

\bibitem{Martinazzo2005b}
{\sc R.~Martinazzo} and {\sc G.~F. Tantardini},
\newblock {\em J. Phys. Chem. A} {\bf 109}, 9379 (2005).

\bibitem{Martinazzo2006a}
{\sc R.~Martinazzo} and {\sc G.~F. Tantardini},
\newblock {\em J. Chem. Phys.} {\bf 124}, 124702 (2006).

\bibitem{Martinazzo2006b}
{\sc R.~Martinazzo} and {\sc G.~F. Tantardini},
\newblock {\em J. Chem. Phys.} {\bf 124}, 124703 (2006).

\bibitem{Morisset2004}
{\sc S.~Morisset}, {\sc F.~Aguillon}, {\sc M.~Sizun}, and {\sc V.~Sidis},
\newblock {\em J. Chem. Phys.} {\bf 121}, 6493 (2004).

\bibitem{Morisset2005}
{\sc S.~Morisset}, {\sc F.~Aguillon}, {\sc M.~Sizun}, and {\sc V.~Sidis},
\newblock {\em J. Chem. Phys.} {\bf 122}, 194702 (2005).

\bibitem{Jackson&Lemoine2001}
{\sc B.~Jackson} and {\sc D.~Lemoine},
\newblock {\em J. Chem. Phys.} {\bf 144}, 474 (2001).

\bibitem{lowe09}
{\sc S.~Casolo}, {\sc R.~Martinazzo}, {\sc M.~Bonfanti}, and {\sc G.~F.
  Tantardini},
\newblock {\em J. Phys. Chem. A} {\bf 113}, 14545 (2009).

\bibitem{Mizes1989}
{\sc H.~A. Mizes} and {\sc J.~S. Foster},
\newblock {\em Science} {\bf 244}, 559 (1989).

\bibitem{Ruffieaux2000}
{\sc R.~Ruffieux}, {\sc O.~Gr{\"o}ning}, {\sc P.~Schwaller}, and {\sc
  L.~Schlapbach},
\newblock {\em Phys. Rev. Lett.} {\bf 84}, 4910 (2000).

\bibitem{Ugeda10}
{\sc M.~M. Ugeda}, {\sc I.~Brihuega}, {\sc F.~Guinea}, and {\sc J.~M.
  G\'omez-Rodr\'\i{}guez},
\newblock {\em Phys. Rev. Lett.} {\bf 104}, 096804 (2010).

\bibitem{TheoInui}
{\sc M.~Inui}, {\sc S.~A. Trugman}, and {\sc E.~Abrahams},
\newblock {\em Phys. Rev. B} {\bf 49}, 3190 (1994).

\bibitem{DewarBook}
{\sc M.~J.~S. Dewar},
\newblock {\em The Molecular Orbital theory of organic chemistry},
\newblock McGraw-Hill, 1969.

\bibitem{Pereira2006}
{\sc V.~M. Pereira}, {\sc F.~Guinea}, {\sc J.~M.~B. {Lopes dos Santos}}, {\sc
  N.~M.~R. Peres}, and {\sc A.~H. {Castro Neto}},
\newblock {\em Phys. Rev. Lett.} {\bf 96}, 036801 (2006).

\bibitem{pereira08a}
{\sc V.~M. Pereira}, {\sc J.~M.~B. {Lopes dos Santos}}, and {\sc A.~H. {Castro
  Neto}},
\newblock {\em Phys. Rev. B} {\bf 77}, 115109 (2008).

\bibitem{Note9}
The approach used is intrinsically periodic. Therefore, the results are best
  viewed as referring to defects which are periodically arranged on
  superlattices with large unit cells.

\bibitem{Boukhvalov2008}
{\sc D.~W. Boukhvalov}, {\sc M.~I. Katsnelson}, and {\sc A.~I. Lichtenstein},
\newblock {\em Phys. Rev. B} {\bf 77}, 035427 (2008).

\bibitem{Note10}
This amounts to consider Kekul\'e structures only, which are much fewer than
  the whole set of $f_S^N$ couplings for all but small $N$ values.

\bibitem{Note11}
It can also be turned into a \protect \emph {charge-density} by
  addition/removal of one electron.

\bibitem{Hornekaer2006II}
{\sc L.~Hornek{\ae}r}, {\sc E.~Rauls}, {\sc W.~Xu}, {\sc
  {\v{Z}}.~{\v{S}}ljivancanin}, {\sc R.~Otero}, {\sc I.~Stensgaard}, {\sc
  E.~Laegsgaard}, {\sc B.~Hammer}, and {\sc F.~Besenbacher},
\newblock {\em Phys. Rev. Lett.} {\bf 97}, 186102 (2006).

\bibitem{ghaderi}
{\sc N.~Ghaderi},
\newblock Private communication, 2010.

\bibitem{ClarGlobet}
{\sc S.~Pogodin} and {\sc I.~Agranat},
\newblock {\em J. Org. Chem.} {\bf 68}, 2720 (2003).

\bibitem{ChemGraphTheo}
{\sc D.~Bonchev} and {\sc D.~H. Rouvray}, editors,
\newblock {\em Chemical Graph Theory}, chapter~6,
\newblock Gordon and Breach, 1991.

\bibitem{RandicChemRev}
{\sc M.~Randi\'c},
\newblock {\em Chem. Rev.} {\bf 103}, 3449 (2003).

\bibitem{LonguetHiggins}
{\sc H.~C. Longuet-Higgins},
\newblock {\em J. Chem. Phys.} {\bf 18}, 265 (1950).

\bibitem{TheoLieb}
{\sc E.~H. Lieb},
\newblock {\em Phys. Rev. Lett.} {\bf 62}, 1201 (1989).

\bibitem{Note12}
Notice that for two electrons in different zero energy state, first-order
  perturbation theory always gives a triplet ground-state. The nature of the
  midgap states and the ensuing interactions with the doubly occupied orbitals
  play a decisive role.

\bibitem{Ovchinnikov}
{\sc A.~A. Ovchinnikov},
\newblock {\em Theoret. Chim. Acta} {\bf 47}, 297 (1978).

\bibitem{Lopez2009}
{\sc M.~P. L\'opez-Sancho}, {\sc F.~de~Juan}, and {\sc M.~A.~H. Vozmediano},
\newblock {\em Phys. Rev. B} {\bf 79}, 075413 (2009).

\bibitem{Rogeau2006}
{\sc N.~Rogeau}, {\sc D.~{Teillet-Billy}}, and {\sc V.~Sidis},
\newblock {\em Chem. Phys. Lett.} {\bf 431}, 135 (2006).

\bibitem{Note13}
Surface puckering upon adsorption, to a good approximation, involves nearest
  neighbouring C atoms only.

\bibitem{Hornekaer2006}
{\sc L.~Hornek{\ae}r}, {\sc {\v{Z}}.~Sljivan{\v{c}}anin}, {\sc W.~Xu}, {\sc
  R.~Otero}, {\sc E.~Rauls}, {\sc I.~Stensgaard}, {\sc E.~L{\ae}gsgaard}, {\sc
  B.~Hammer}, and {\sc F.~Besenbacher},
\newblock {\em Phys. Rev. Lett.} {\bf 96}, 156104 (2006).

\bibitem{Note14}
There is a long open search for efficient pathways leading to $H_2$ formation
  on graphitic surfaces, because of its importance in explaining the observed
  abundance of molecular hydrogen in the interstellar medium.

\bibitem{Note15}
Remember that the relaxation energy for the single H atom ($\sim 0.8$ eV) has
  the same magnitude as the overall binding energy, \protect \emph {i.e.} about
  half of the bond formation energy is spent for relaxing the substrate.

\bibitem{Hornekaer2007}
{\sc L.~Hornek{\ae}r}, {\sc W.~Xu}, {\sc R.~Otero}, {\sc E.~L{\ae}gsgaard}, and
  {\sc F.~Besenbacher},
\newblock {\em Chem. Phys. Lett.} {\bf 446}, 237 (2007).

\bibitem{Ferro09}
{\sc Y.~Ferro}, {\sc S.~Morisset}, and {\sc A.~Allouche},
\newblock {\em Chem. Phys. Lett.} {\bf 478}, 42 (2009).

\bibitem{sofo07}
{\sc J.~O. Sofo}, {\sc A.~S. Chaudhari}, and {\sc G.~D. Barber},
\newblock {\em Phys. Rev. B} {\bf 75}, 153401 (2007).

\bibitem{Lebegue09}
{\sc S.~Leb\`egue}, {\sc M.~Klintenberg}, {\sc O.~Eriksson}, and {\sc M.~I.
  Katsnelson},
\newblock {\em Phys. Rev. B} {\bf 79}, 245117 (2009).

\bibitem{Elias09}
{\sc D.~C. Elias}, {\sc R.~R. Nair}, {\sc T.~M.~G. Mohiuddin}, {\sc S.~V.
  Morozov}, {\sc P.~Blake}, {\sc M.~P. Halsall}, {\sc A.~C. Ferrari}, {\sc
  D.~W. Boukhvalov}, {\sc M.~I. Katsnelson}, {\sc A.~K. Geim}, and {\sc K.~S.
  Novoselov},
\newblock {\em Science} , 5914 (2009).

\bibitem{casoloTesi}
{\sc S.~Casolo},
\newblock {\em Adsorption, Clustering and reaction of H atoms in graphene},
\newblock PhD thesis, University of Milan, 2009.

\bibitem{flores09}
{\sc M.~Z.~S. Flores}, {\sc P.~A.~S. Autreto}, {\sc S.~B. Legoas}, and {\sc
  D.~S. Galvao},
\newblock {\em Nanotechnology} {\bf 20}, 465704 (2009).

\bibitem{Hornekaer2009}
{\sc R.~Balog}, {\sc B.~J{\o}rgensen}, {\sc J.~Wells}, {\sc E.~L{\ae}gsgaard},
  {\sc P.~Hofmann}, {\sc F.~Besenbacher}, and {\sc L.~Hornek{\ae}r},
\newblock {\em J. Am. Chem. Soc.} {\bf 25}, 131 (2009).

\bibitem{Luo09}
{\sc Z.~Luo}, {\sc T.~Yu}, {\sc K.~Kim}, {\sc Y.~You}, {\sc S.~Lim}, {\sc
  Z.~Shen}, {\sc S.~Wang}, and {\sc J.~Lin},
\newblock {\em ACS Nano} {\bf 3}, 1781 (2009).

\bibitem{Brus09}
{\sc N.~Jung}, {\sc N.~Kim}, {\sc S.~Jockusch}, {\sc N.~J. Turro}, {\sc
  P.~Kim}, and {\sc L.~Brus},
\newblock {\em Nano Lett.} {\bf 8}, 4597 (2009).

\bibitem{Ruffieux2002}
{\sc P.~Ruffieux}, {\sc O.~Gr{\"o}ning}, {\sc M.~Bielmann}, {\sc P.~Mauron},
  {\sc L.~Schlapbach}, and {\sc P.~Gr{\"o}ning},
\newblock {\em Phys. Rev. Lett.} {\bf 66}, 245416 (2002).

\bibitem{Ruffieux02}
{\sc P.~Ruffieux}, {\sc O.~Gr{\"o}ning}, {\sc M.~Bielmann}, {\sc P.~Mauron},
  {\sc L.~Schlapbach}, and {\sc P.~Gr{\"o}ning},
\newblock {\em Phys. Rev. B} {\bf 66}, 245416 (2002).

\bibitem{Schwierz2010}
{\sc F.~Schwierz},
\newblock {\em Nature Nanothech.} {\bf 5}, 487 (2010).

\bibitem{avouris10}
{\sc Y.-M. Lin}, {\sc C.~Dimitrakopoulos}, {\sc K.~A. Jenkins}, {\sc D.~B.
  Farmer}, {\sc H.-Y. Chiu}, {\sc A.~Grill}, and {\sc P.~Avouris},
\newblock {\em Science} {\bf 327}, 662 (2010).

\bibitem{Avouris07}
{\sc P.~Avouris}, {\sc Z.~Chen}, and {\sc V.~Perebeinos},
\newblock {\em Nat. Nanotech.} {\bf 2}, 605 (2007).

\bibitem{Note16}
The defects discussed in this chapter, along with charged scatterer, are
  ascending as the most likely origin of the conductivity minimum \protect
  \citep {peres10}. The counter-intuitive role of defects in \protect \emph
  {increasing} the conductivity finds its origin in the modification of the
  graphene DOS close to the Dirac point.

\bibitem{zhou07}
{\sc S.~Y. Zhou}, {\sc G.-H. Gweon}, {\sc A.~V. Fedorov}, {\sc P.~N. First},
  {\sc W.~A. de~Heer}, {\sc D.-H. Lee}, {\sc F.~Guinea}, {\sc A.~H. {Castro
  Neto}}, and {\sc A.~Lanzara},
\newblock {\em Nat. Mater.} {\bf 6}, 770 (2007).

\bibitem{Bostwick07}
{\sc A.~Bostwick}, {\sc T.~Ohta}, {\sc T.~Seyller}, {\sc K.~Horn}, and {\sc
  E.~Rotenberg},
\newblock {\em Nat. Phys.} {\bf 3}, 36 (2007).

\bibitem{martinazzo10}
{\sc R.~Martinazzo}, {\sc S.~Casolo}, and {\sc G.~F. Tantardini},
\newblock {\em Phys. Rev. B} {\bf 81}, 245420 (2010).

\bibitem{Note17}
As shown in the previous sections one can equivalently introduce either a
  vacancy or an adatom.

\bibitem{louie06}
{\sc Y.-W. Son}, {\sc M.~L. Cohen}, and {\sc S.~G. Louie},
\newblock {\em Phys. Rev. Lett.} {\bf 97}, 216803 (2006).

\bibitem{Antidot1}
{\sc T.~G. Pedersen}, {\sc C.~Flindt}, {\sc J.~Pedersen}, {\sc N.~A.
  Mortensen}, {\sc A.~Jauho}, and {\sc K.~Pedersen},
\newblock {\em Phys. Rev. Lett.} {\bf 100}, 136804 (2008).

\bibitem{Antidot2}
{\sc J.~A. F\"urst}, {\sc T.~G. Pedersen}, {\sc M.~Brandbyge}, and {\sc A.-P.
  Jauho},
\newblock {\em Phys. Rev. B} {\bf 80}, 115117 (2009).

\bibitem{Antidot3}
{\sc W.~Liu}, {\sc Z.~F. Wang}, {\sc Q.~W. Shi}, {\sc J.~Yang}, and {\sc
  F.~Liu},
\newblock {\em Phys. Rev. B} {\bf 80}, 233405 (2009).

\bibitem{meyer08}
{\sc J.~C. Meyer}, {\sc C.~O. Girit}, {\sc M.~F. Crommie}, and {\sc A.~Zettl},
\newblock {\em Appl. Phys. Lett.} {\bf 92}, 123110 (2008).

\bibitem{fischbein08}
{\sc M.~D. Fischbein} and {\sc M.~Drndic},
\newblock {\em Appl. Phys. Lett.} {\bf 93}, 113107 (2008).

\bibitem{antidot08}
{\sc T.~Shen}, {\sc Y.~Q. Wu}, {\sc M.~A. Capano}, {\sc L.~P. Rokhinson}, {\sc
  L.~W. Engel}, and {\sc P.~D. Ye},
\newblock {\em Appl. Phys. Lett.} {\bf 93}, 122102 (2008).

\bibitem{weiss09}
{\sc J.~Eroms} and {\sc D.~Weiss},
\newblock {\em New J. Phys.} {\bf 11}, 095021 (2009).

\bibitem{GrapheneHoles}
{\sc J.~W. Bai}, {\sc X.~Zhong}, {\sc S.~Jiang}, {\sc Y.~H. Y}, and {\sc X.~F.
  Duan},
\newblock {\em Nature Nanotech.} {\bf 5}, 190 (2010).

\bibitem{HornekaerNature}
{\sc R.~Balog}, {\sc B.~Jorgensen}, {\sc L.~Nilsson}, {\sc M.~Andersen}, {\sc
  E.~Rienks}, {\sc M.~Bianchi}, {\sc M.~Fanetti}, {\sc E.~Laegsgaard}, {\sc
  A.~Baraldi}, {\sc S.~Lizzit}, {\sc Z.~Sljivancanin}, {\sc F.~Besenbacher},
  {\sc B.~H. B}, {\sc T.~G. Pedersen}, {\sc P.~Hofmann}, and {\sc
  L.~Hornek{\ae}r},
\newblock {\em Nat. Mater.} {\bf 4}, 315 (2010).

\bibitem{Note18}
The number of $E$ irreps is always even because of the `coalescence' of the two
  valleys. Thus, one only needs to remove the accidental degeneracy created by
  such folding.

\bibitem{Panchakarla09}
{\sc L.~S. Panchakarla}, {\sc K.~S. Subrahmanyam}, {\sc S.~Saha}, {\sc
  A.~Govindaraj}, {\sc H.~R. Krishnamurthy}, {\sc U.~V. Wagmre}, and {\sc
  C.~N.~R. Rao},
\newblock {\em Adv. Mater.} {\bf 21}, 4726 (2009).

\bibitem{ciNature10}
{\sc L.~Ci}, {\sc L.~Song}, {\sc C.~Jin}, {\sc D.~Jariwala}, {\sc D.~Wu}, {\sc
  Y.~Li}, {\sc A.~Srivastava}, {\sc Z.~F. Wang}, {\sc K.~Storr}, {\sc
  L.~Balicas}, {\sc F.~Liu}, and {\sc P.~M. Ajayan},
\newblock {\em Nature Mater.} {\bf 9}, 430 (2010).

\bibitem{Pontes09}
{\sc R.~B. Pontes}, {\sc A.~Fazzio}, and {\sc G.~M. Dalpian},
\newblock {\em Phys. Rev. B} {\bf 79}, 033412 (2009).

\bibitem{casolo10II}
{\sc S.~Casolo}, {\sc R.~Martinazzo}, and {\sc G.~F. Tantardini},
\newblock {\em J. Phys. Chem. C} {\bf 115}, 3250 (2011).

\end{thebibliography}

\end{document}